\documentclass[a4paper,10pt]{article}
\usepackage[round]{natbib}
\usepackage{enumerate}
\usepackage{times,amsmath,graphicx}
\usepackage{color}
\usepackage{multirow}
\usepackage[flushleft]{threeparttable}
\usepackage{bm}
\usepackage{geometry}
\usepackage{nohyperref}
\usepackage{url}
\usepackage[symbol]{footmisc}

\usepackage{etoolbox}
\makeatletter
\patchcmd{\@verbatim}
  {\verbatim@font}
  {\verbatim@font\small}
  {}{}
\makeatother
\usepackage{listings}
\usepackage{titlesec}
\titleformat*{\section}{\large\bfseries}
\titleformat*{\subsection}{\large\bfseries}
\titleformat*{\subsubsection}{\normalsize\bfseries}

\begin{document}
\title{\huge \textbf{Bayesian survival analysis with BUGS}}

\author{{\large Danilo Alvares$^{1}$\footnote{E-mail: dalvares@mat.uc.cl}\ , \ Elena L{\'a}zaro$^{2}$, \ Virgilio G{\'o}mez-Rubio$^{3}$, \ and \ Carmen Armero$^{4}$} \\ \\ \small $^{1}$Department of Statistics, Pontificia Universidad Cat{\'o}lica de Chile, Macul, Santiago, Chile \\ \small $^{2}$Plant Protection and Biotechnology Centre, Instituto Valenciano de Investigaciones Agrarias, Moncada, Spain \\ \small $^{3}$Department of Mathematics, School of Industrial Engineering, Universidad de Castilla-La Mancha, Albacete, Spain \\ \small $^{4}$Department of Statistics and Operational Research, Universitat de Val{\`e}ncia, Burjassot, Spain \\ 
}

\date{}

\maketitle

\begin{abstract}
\noindent Survival analysis is one of the most important fields of statistics in medicine and biological sciences. In addition, the computational advances in the last decades have favoured the use of Bayesian methods in this context, providing a flexible and powerful alternative to the traditional frequentist approach. The objective of this paper is to summarise some of the most popular Bayesian survival models, such as accelerated failure time, proportional hazards, mixture cure, competing risks, multi-state, frailty, and joint models of longitudinal and survival data. Moreover, an implementation of each presented model is provided using a BUGS syntax that can be run with JAGS from the R programming language. Reference to other Bayesian R-packages are also discussed. \vspace{0.1cm}

\end{abstract}

\section{Introduction} \label{sec1}

Survival analysis, sometimes referred to as \textit{failure-time analysis}, is one of the most important fields of statistics, mainly in medicine and biological sciences \citep{kleinbaum2012,collet2015}. Survival times are data that measure follow-up time from a defined starting point to the occurrence of a given event of interest or endpoint, for instance, onset of disease, cure, death, etc. \citep{kalbfleisch2002}.

Continuous survival times are defined as non-negative random variables, $T$, whose probabilistic behaviour can be equivalently described by its \textsl{hazard function}, \textit{survival function}, or \textit{density function}. The hazard function of $T$ at time $t$ represents the instantaneous rate of the event occurrence for the population group that is still at risk at time $t$. It is defined as
\begin{equation}
h(t \mid \bm \theta) = \lim_{\Delta t \to 0} \frac{P(t \leq T < t+\Delta t \mid T \geq t, \bm \theta)}{\Delta t} = \frac{f(t \mid \bm \theta)}{S(t \mid \bm \theta)}, \;\;\; t>0, \label{relation1}
\end{equation}

\noindent where $\boldsymbol \theta$ is a set of unknown quantities, $f(t \mid \bm \theta)$ is the density function of $T$ given $\boldsymbol \theta$, and $S(t \mid \bm \theta)=P(T > t \mid \bm \theta)$ is the survival function of $T$ given $\boldsymbol \theta$. Note that $h(t \mid \bm \theta) \geq 0$ and in general $\int h(t \mid \bm \theta)\,\mbox{d}t=\infty$. The hazard function is defined in terms of a conditional probability but is not a probability. It could be interpreted as the approximate instantaneous probability of having the event given that the target individual has not yet experienced it at time $t$.

Usual relationships between $h(t \mid \bm \theta)$, $f(t \mid \bm \theta)$, and $S(t \mid \bm \theta)$ derived from (\ref{relation1}) that will be useful throughout the article include
\begin{eqnarray}
H(t \mid  \bm \theta) &=& \int_{0}^t \, h(u \mid \boldsymbol \theta)\, \mbox{d}u, \;\;\; t > 0, \label{eq:cumhazard} \\
S(t \mid  \bm \theta) &=& \exp\big\{-H(t \mid \boldsymbol \theta)\big\}, \;\;\; t > 0, \label{eq:survival} \\
f(t \mid \bm \theta)  &=& h(t \mid \bm \theta)\,\exp\big\{-H(t \mid \boldsymbol \theta)\big\}, \;\;\; t > 0, \label{eq:density}
\end{eqnarray}

\noindent where $H(t \mid \bm \theta)$ denotes the \textit{cumulative hazard function} of $T$.

Standard statistical techniques cannot usually be applied to survival data because in most cases they are not fully recorded, mainly due to censoring and/or truncation schemes \citep{klein2013}. Also, normal distributions are usually inappropriate for analyzing survival data at their original scale, since they are positive and often asymmetrical. Distributions such as the Weibull, gamma or log-normal are the ones that now occupy the leading role.

From a Bayesian perspective, censoring mechanisms do not affect the prior distribution but their do affect the likelihood function, which factorises in general as the product of the likelihood function corresponding to each individual $i$ in the sample as $L(\bm \theta)=\prod_{i=1}^{n}\, L_{i}(\bm \theta)$. In general terms, censored observations can be classified as \textit{right-censored}, \textit{left-censored} and \textit{interval-censored}. See \cite{klein2003} for technical details, interpretations and examples. The contribution of the survival observation $t$ of individual $i$ to the likelihood function, $L_{i}(\bm \theta)$, depends on the type of censoring as follows:
\begin{center}
\begin{tabular}{ll}
Exact survival time:           & $f_i(t\mid \bm \theta)$ \\
Right-censored observation:    & $S_i(C_r \mid \bm \theta)$ \\
Left-censored observation:     & $1-S_i(C_l \mid \bm \theta)$ \\
Interval-censored observation: & $S_i(C_l \mid \bm \theta) - S_i(C_r \mid \bm \theta)$
\end{tabular}
\end{center} \vspace{0.15cm}

Right censoring here is type I, where the end of study period $C_r$ is known and prefixed before it begins. In the case of left-censored observations, the survival time is known to be below a certain censoring time $C_l$. When survival times are interval-censored in $[C_l, \,C_r]$, the survival time is somewhere in this interval.

Due to these peculiar characteristics and their extreme relevance to different scientific subjects, survival models/methods have been extensively developed over the past 50 years, which includes many R-packages\footnote{\url{https://cran.r-project.org/web/views/Survival.html}}. Hence, in order to illustrate for which situations some survival models are more appropriate than others, we summarise the main models for survival data, such as accelerated failure time (AFT), proportional hazards (PH), mixture cure, competing risks, multi-state, frailty, and joint models of longitudinal and survival data. In particular, the inferential process is performed from a Bayesian perspective, which is an appealing alternative to the traditional frequentist approach, since the Bayesian probability conception allows to measure the uncertainty associated with parameters, models, hypotheses, missing data in probabilistic terms \citep{ibrahim2001}. So, based on the Bayesian paradigm, we exemplify each survival modelling using Markov chain Monte Carlo (MCMC) methods \citep{bda2013} implemented in \texttt{BUGS} syntax \citep{gilks1994} with the support of \texttt{JAGS} \citep{plummer2003} and the \texttt{rjags} \citep{rjags} package (version 4-10) for the R language (version 4.0.2) \citep{R}.

The set of models introduced in this paper is quite comprehensive as it covers the main models in survival analysis. However, there are a number of R-packages to fit different types of Bayesian survival models which can be used in specific contexts. For example, if one is interested in AFT modelling, the \texttt{bayesSurv} package provides implementations of mixed-effects AFT models with various censored data specifications \citep{bayesSurv}; while the \texttt{DPpackage} package fits nonparametric and semi-parametric models with different prior processes \citep{DPpackage}; and \texttt{spBayesSurv} package includes AFT and PH models (among others) for spatial/non-spatial survival data \citep{spBayesSurv}. If a PH specification is preferable, the \texttt{BMA} package implements a Bayesian model averaging approach for Cox PH models \citep{BMA}; the \texttt{dynsurv} package provides time-varying coefficient models for interval-censored and right-censored survival data \citep{dynsurv}; \texttt{ICBayes} package offers semi-parametric regression survival models to interval-censored time-to-event data \citep{ICBayes}; and the \texttt{spatsurv} package fits parametric PH spatial survival models \citep{spatsurv}. For multi-state problems, we highlight the \texttt{CFC} package that implements a cause-specific framework for competing-risk analysis \citep{CFC}; and \texttt{SemiCompRisks} package that provides a broader specification using of independent/clustered semi-competing risks models \citep{SemiCompRisks}. When a joint approach of longitudinal and survival data is required, the most popular package is the \texttt{JMbayes} one, where the longitudinal process is modelled through a linear mixed framework, a Cox PH model is assumed for the survival process, and various association structures between both processes are provided \citep{JMbayes}. Other generic packages for Bayesian inference such as, for example, \texttt{BayesX} \citep{BayesX}, \texttt{RStan} \citep{Stan,RStan} or \texttt{INLA} \citep{INLA} can be used to fit different survival models.

The paper is organised as follows. Sections~\ref{sec:survreg}--\ref{sec:JM} describe the following survival models: AFT, PH, mixture cure, competing risks, multi-state, frailty, and joint models of longitudinal and survival data. In particular, in each one of these sections, we briefly introduce notation and basic concepts of the models under discussion. Then, the survival dataset used (available in an R-package) is described jointly with the \texttt{BUGS} code implementation. Finally, posterior summaries, and graphs of quantities of interest derived from the posterior distribution are provided. To conclude, Section~\ref{sec:conclusions} ends with an overview of the Bayesian survival models introduced and implemented in this paper, motivating the use and adaptation of the codes provided. Theoretical and methodological aspects of survival models and Bayesian inference are not discussed in depth in this paper, so we recommend that readers unfamiliar with these topics review specific references, such as \cite{klein2013} and \cite{bda2013}.

\section{Survival regression models} \label{sec:survreg}

Regression models focus on the association between time-to-event random variables and covariates (explanatory/predictor variables or risk factors). They allow the comparison of survival times between groups both with regard to the general behaviour of the individuals in each group and prediction for new members of them. The most popular approaches to survival regression models are the accelerated failure time (AFT) and the proportional hazards (PH) models, also known as Cox models \citep{cox1972}.

\subsection{Accelerated failure time models} \label{sec:AFT}

AFT models are the survival counterpart of linear models. Survival time $T$ in logarithmic scale is expressed in terms of a linear combination of covariates ${\bm x}$ with regression coefficients $\bm \beta$ and a measurement error $\xi$ as follows:
\begin{equation}
\log\,(T)= {\bm x}^{\top}{\bm \beta} + \sigma \,\epsilon, \label{aftmodel}
\end{equation}

\noindent where $\sigma$ is a scale parameter and $\epsilon$ an error term usually expressed via a normal, logistic or a standard Gumbel probabilistic distribution. The particular case of a standard Gumbel distribution for $\epsilon$ implies a conditional (on $\bm \beta$ and $\sigma$) Weibull survival model for $T$ with shape $\alpha = 1/\sigma$ and scale $\lambda({\bm \beta}, \sigma) = \exp\left\{-{\bm x}^{\top}{\bm \beta}/\sigma\right\}$ parameters \citep{christensen2011}.

\subsubsection{\textit{larynx} dataset}

We consider a larynx cancer dataset, referred to as \textit{larynx}. It is available from the \texttt{KMsurv} package \citep{KMsurv}:
\begin{verbatim}
R> library("KMsurv")
R> data("larynx")
R> str(larynx)
'data.frame':	90 obs. of  5 variables:
 $ stage : int  1 1 1 1 1 1 1 1 1 1 ...
 $ time  : num  0.6 1.3 2.4 2.5 3.2 3.2 3.3 3.3 3.5 3.5 ...
 $ age   : int  77 53 45 57 58 51 76 63 43 60 ...
 $ diagyr: int  76 71 71 78 74 77 74 77 71 73 ...
 $ delta : int  1 1 1 0 1 0 1 0 1 1 ...
\end{verbatim}

This dataset provides observations of 90 male larynx-cancer patients, diagnosed and treated in the period 1970--1978 \citep{kardaun1983}. The following variables were observed for each patient:
\begin{itemize}
	\item \texttt{stage}: disease stage (1--4).
	\item \texttt{time}: time (in months) from first treatment until death, or end of study.
	\item \texttt{age}: age (in years) at diagnosis of larynx cancer.
	\item \texttt{diagyr}: year of diagnosis of larynx cancer.
	\item \texttt{delta}: death indicator (1: if patient died; 0: otherwise).
\end{itemize}

\subsubsection{Model specification}

Survival times are analysed through the following accelerated failure time (AFT) model:
\begin{equation}
\log(T) = \beta_{1} + \beta_{2}\mbox{I}_{(\texttt{stage}=2)} + \beta_{3}\mbox{I}_{(\texttt{stage}=3)} + \beta_{4}\mbox{I}_{(\texttt{stage}=4)} + \beta_{5}\texttt{age} +\beta_{6}\texttt{diagyr} + \sigma \epsilon, \label{eq:larynx1}
\end{equation}

\noindent where $T$ represents death time for each individual; $\beta_{1}$ is an intercept; $\mbox{I}_{(\texttt{stage}=\cdot)}$ is an indicator variable for \texttt{stage}=2, 3, 4 with regression coefficients $\beta_{2}$, $\beta_{3}$, and $\beta_{4}$, respectively (\texttt{stage}=1 is considered as the reference category); and $\beta_{5}$ and $\beta_{6}$ are regression coefficients for \texttt{age} and \texttt{diagyr} covariates, respectively. The errors $\epsilon$'s are i.i.d. random variables which follow a standard Gumbel distribution and $\sigma$ is a scale parameter.

The Bayesian model is completed with the specification of a prior distribution for their corresponding parameters. A non-informative prior independent default scenario is considered. The marginal prior distribution for each regression coefficient $\beta_{k}$, $k=1,\ldots,6$, is elicited as a normal distribution centered at zero and with a small precision, N($0,0.001$). A uniform distribution, Un($0,10$), is selected as the marginal prior distribution for $\alpha$ (i.e., the Weibull shape parameter). However, the use of another continuous distributions, e.g., Gamma($1,0.001$) is also common under a non-informative prior framework \cite{sahu1997}.

\subsubsection{Model implementation}

Censoring is handled in \texttt{JAGS} using the distribution \texttt{dinterval} which implies the modification of the default \texttt{time} variable and the creation of two new ones, \texttt{cens} and \texttt{is.censored}. For a generic database which deals with observations of different censoring types (i.e., right-censored, left-censored, and/or interval-censored) the specification of these variables should be managed as follows:
\begin{itemize}
  \item \texttt{time}: survival time is \texttt{time} if the event was observed (\texttt{delta}=1); otherwise, \texttt{NA}.
  \item \texttt{cens[i,]}: two column matrix with $n$ rows (with $n$ equal to the number of individuals). \texttt{cens[i,]} aggregates for each observation $i$ a vector of two cut points. According to the censoring type \texttt{cens} must be specified as:
  \begin{itemize}
			\item Uncensored : \texttt{cens[i,1] = 0} \, (or any arbitrary value is allowed).
			\item Right-censored: \texttt{cens[i,1] =} \, censoring time.
			\item Left-censored: \texttt{cens[i,2] =} \, censoring time.
			\item Interval-censored: \texttt{cens[i,] =} \, (cens.low, \ cens.up) \; (cens.low corresponds with the lower limit and cens.up with the upper limit of the observed interval).
  \end{itemize}
  \item \texttt{is.censored}: a help ordinal variable to indicate the censoring status. According to the censoring type \texttt{cens} must be specified as:
  \begin{itemize}
			\item Uncensored: \texttt{is.censored = 0}
			\item Right-censored: \texttt{is.censored = 1}
			\item Left-censored: \texttt{is.censored = 0}
			\item Interval-censored: \texttt{is.censored = 1}
  \end{itemize}
\end{itemize}

Thus, for the \texttt{larynx} dataset which contains uncensored and right-censored observations the following code is dedicated to creating or transforming these variables:
\begin{verbatim}
R> # Survival and censoring times
R> cens <- matrix(c(larynx$time, rep(NA, length(larynx$time))),
+    nrow = length(larynx$time), ncol = 2)
R> larynx$time[larynx$delta == 0] <- NA # Censored
R> is.censored <- as.numeric(is.na(larynx$time))
\end{verbatim}

Next, we have scaled the \texttt{age} and \texttt{diagyr}, and encoded the \texttt{stage} as a factor. Then, we have created a design matrix \texttt{X} of these covariates:
\begin{verbatim}
R> larynx$age <- as.numeric(scale(larynx$age))
R> larynx$diagyr <- as.numeric(scale(larynx$diagyr))
R> larynx$stage <- as.factor(larynx$stage)
R> X <- model.matrix(~ stage + age + diagyr, data = larynx)
\end{verbatim}

Listing~\ref{lst:aft} shows a generic implementation of an AFT model in \texttt{BUGS} syntax using \textit{larynx} data.

\begin{figure}[htb!] \small
\begin{lstlisting}[caption={AFT model in \texttt{BUGS} syntax (file named as \textbf{AFT.txt}).}, label={lst:aft}]
model{
  for(i in 1:n){
    # Survival and censoring times
    is.censored[i] ~ dinterval(time[i],cens[i,1])
    time[i] ~ dweib(alpha,lambda[i])
    lambda[i] <- exp(-mu[i] * alpha)
    mu[i] <- inprod(beta[],X[i,])
  }

  # Prior distributions
  for(l in 1:Nbetas){ beta[l] ~ dnorm(0,0.001) }
  alpha ~ dunif(0,10)
}
\end{lstlisting}
\end{figure}

Once the \texttt{BUGS} syntax and its corresponding variables has been created, \texttt{JAGS} requires specifying some elements to run the MCMC simulation:
\begin{itemize}
    \item \texttt{d.jags}: a list with all the elements/data specified in the model.
    \item \texttt{i.jags}: a function that returns a list of initial random values for each model parameters. 
    \item \texttt{p.jags}: a character vector with the parameter names to be monitored/saved.
\end{itemize}

These elements are defined for our AFT model as follows:
\begin{verbatim}
R> d.jags <- list(n = nrow(larynx), time = larynx$time, cens = cens, X = X,
+    is.censored = is.censored, Nbetas = ncol(X))
R> i.jags <- function(){ list(beta = rnorm(ncol(X)), alpha = runif(1)) }
R> p.jags <- c("beta", "alpha")
\end{verbatim}

Then, MCMC algorithm is run in three steps. Firstly, the \texttt{JAGS} model is compiled by means of the \texttt{jags.model} function available from the \texttt{rjags} package:
\begin{verbatim}
R> library("rjags")
R> m1 <- jags.model(data = d.jags, file = "AFT.txt", inits = i.jags, n.chains = 3)
\end{verbatim}

The \texttt{n.chains} argument indicates the number of Markov chains selected. Secondly, a burn-in period is considered (here the first 1000 simulations) using the \texttt{update} function:
\begin{verbatim}
R> update(m1, 1000)
\end{verbatim}

Thirdly, the model is run using \texttt{coda.samples} function for a specific number of iterations to monitor (here \texttt{n.iter}=50000) as well as a specific thinning value (here \texttt{thin}=10) in order to reduce autocorrelation in the saved samples:
\begin{verbatim}
R> res <- coda.samples(m1, variable.names = p.jags, n.iter = 50000, thin = 10)
\end{verbatim}

A posterior distributions summary can be obtained through the \texttt{summary} function:
\begin{verbatim}
R> summary(res)
Iterations = 2010:52000
Thinning interval = 10 
Number of chains = 3 
Sample size per chain = 5000

1. Empirical mean and standard deviation for each variable,
   plus standard error of the mean:
            Mean     SD Naive SE Time-series SE
alpha    1.03426 0.1353 0.001105       0.001365
beta[1]  2.55344 0.2967 0.002422       0.003868
beta[2] -0.12702 0.4769 0.003894       0.004812
beta[3] -0.64710 0.3628 0.002962       0.004125
beta[4] -1.66226 0.4431 0.003618       0.004878
beta[5] -0.20953 0.1552 0.001267       0.001323
beta[6]  0.07054 0.1622 0.001324       0.001523

2. Quantiles for each variable:
           2.5%      25%      50%     75%    97.5%
alpha    0.7827  0.93886  1.02986  1.1216  1.31162
beta[1]  2.0427  2.34542  2.52919  2.7382  3.18745
beta[2] -1.0342 -0.44200 -0.13807  0.1745  0.86159
beta[3] -1.3800 -0.88079 -0.63669 -0.4061  0.05082
beta[4] -2.5689 -1.94726 -1.65245 -1.3664 -0.82138
beta[5] -0.5279 -0.31022 -0.20582 -0.1040  0.08472
beta[6] -0.2342 -0.04002  0.06568  0.1766  0.40106
\end{verbatim}

Markov chains must reach stationarity and, to check this condition, several convergence diagnostics can be applied. Trace plots (\texttt{traceplot} function) and the calculation of the Gelman-Rubin diagnostic (\texttt{gelman.diag} function) are used for illustrating this issue \citep{gelman1992, brooks1998} with the \texttt{coda} package \citep{coda}, which is already loaded as \texttt{rjags} depends on it:
\begin{verbatim}
R> par(mfrow = c(2,4))
R> traceplot(res)
\end{verbatim}

\begin{verbatim}
R> gelman.diag(res)
Potential scale reduction factors:
        Point est. Upper C.I.
alpha            1       1.00
beta[1]          1       1.01
beta[2]          1       1.00
beta[3]          1       1.01
beta[4]          1       1.00
beta[5]          1       1.00
beta[6]          1       1.00
Multivariate psrf
1
\end{verbatim}

\begin{figure}[bt]
\centering
\includegraphics[scale=0.6]{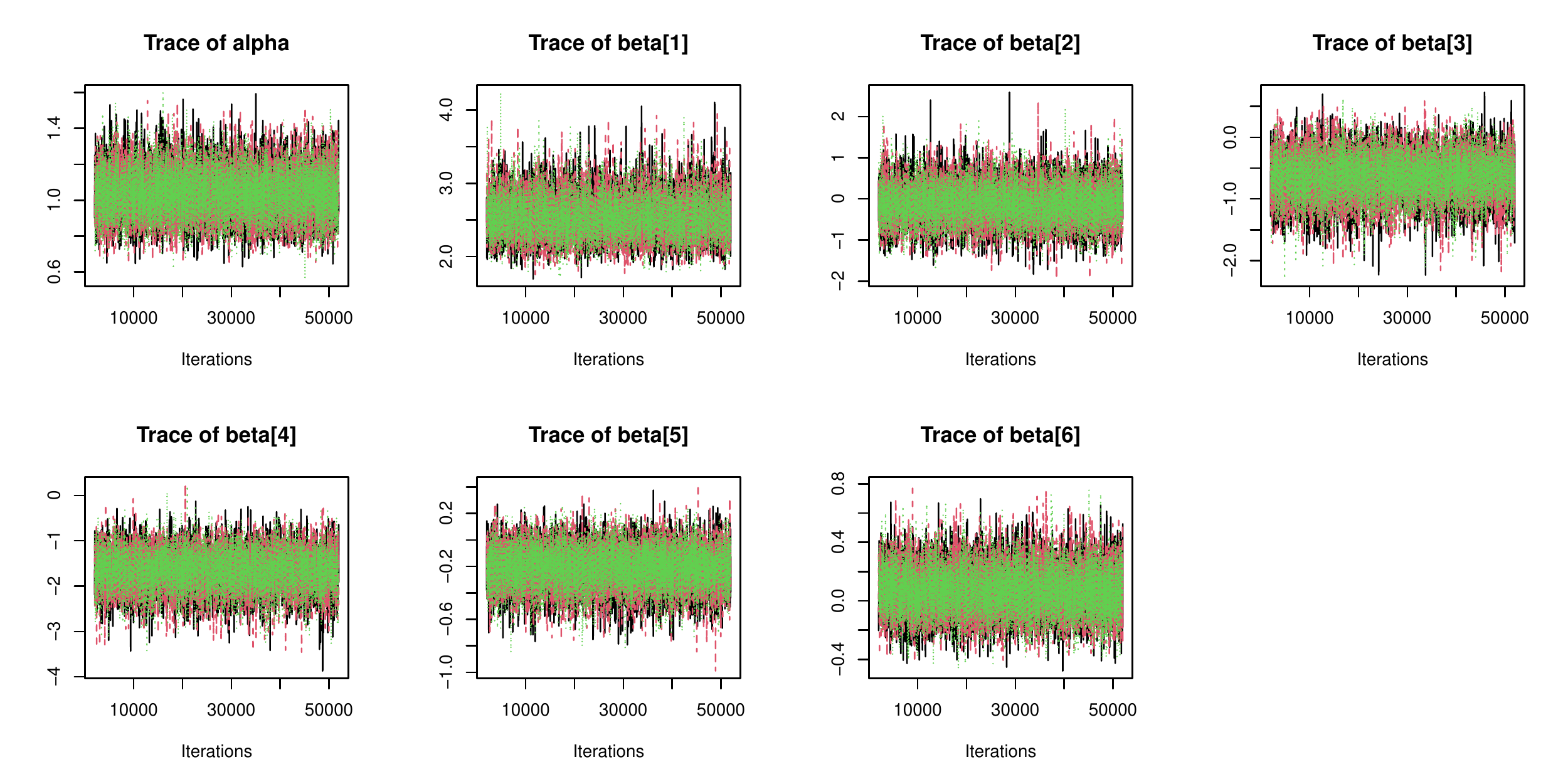}
\caption{\label{fig:AFT1}Trace plots for $\alpha$ and $\beta$'s of the AFT model (\ref{eq:larynx1}).}
\end{figure}

Trace plots of the sampled values for each parameter in the chain appear overlapping one another and Gelman-Rubin values are very close to 1, which indicates that convergence has been achieved.

From the \texttt{densplot} function, also available from the \texttt{coda} package, we can draw the marginal posterior distributions (using kernel smoothing) for all the model parameters:
\begin{verbatim}
R> par(mfrow = c(2,4))
R> densplot(res, xlab = "")
\end{verbatim}

\begin{figure}[bt]
\centering
\includegraphics[scale=0.6]{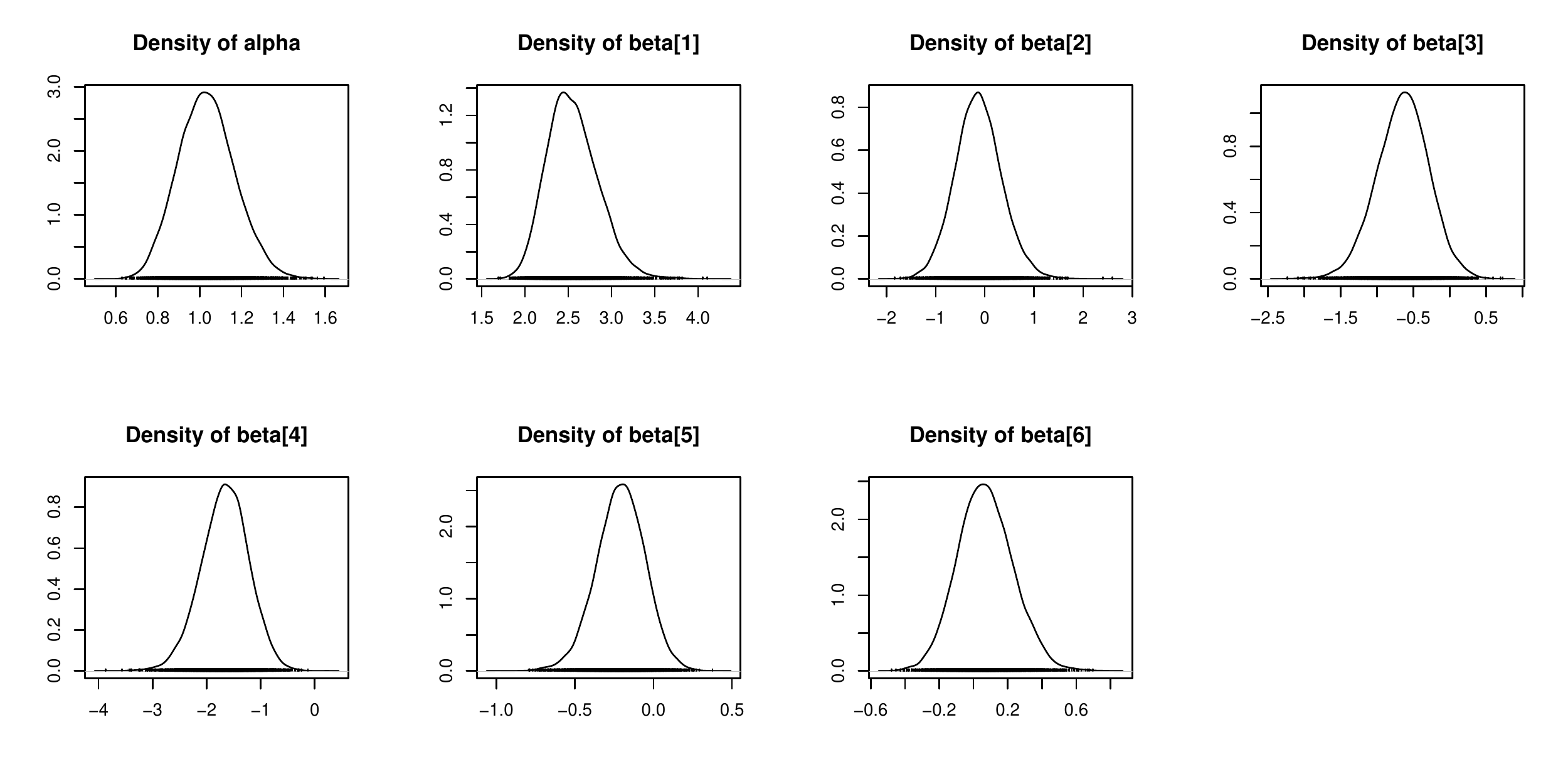}
\caption{\label{fig:AFT2}Density plots for $\alpha$ and $\beta$'s of the AFT model (\ref{eq:larynx1}).}
\end{figure}

Simulation-based Bayesian inference requires using simulated samples to summarise posterior distributions or calculate any relevant quantities of interest. So, posterior samples from the three Markov chains can be merged together using the following code:
\begin{verbatim}
R> result <- as.mcmc(do.call(rbind, res))
\end{verbatim}

Next, the posterior samples of each parameter are extracted from the \texttt{result} object as follows:
\begin{verbatim}
R> alpha <- result[,1]; beta1 <- result[,2]; beta2 <- result[,3]; beta3 <- result[,4]
R> beta4 <- result[,5]; beta5 <- result[,6]; beta6 <- result[,7]
\end{verbatim}

A relevant quantity for AFT models is the \textit{relative median} (RM) between two individuals with covariate vectors ${\bm x}_{1}$ and ${\bm x}_{2}$. This measure compares the median of the survival time between both individuals and is defined as: $$\mbox{RM}({\bm x}_{1},{\bm x}_{2}, {\bm \beta}) = \exp\left\{({\bm x}_{1} - {\bm x}_{2})^{\top}{\bm \beta}\right\}.$$

As an illustration, we can easily summarise the posterior distribution of the RM between two men of the same \texttt{age} and \texttt{diagyr} (year of diagnosis) but one in \texttt{stage}=3 and the other in \texttt{stage}=4:
\begin{verbatim}
R> RM.s3_s4 <- exp(beta3 - beta4)
R> summary(RM.s3_s4)
Iterations = 1:15000
Thinning interval = 1 
Number of chains = 1 
Sample size per chain = 15000 

1. Empirical mean and standard deviation for each variable,
   plus standard error of the mean:
          Mean             SD       Naive SE Time-series SE 
       3.01520        1.34859        0.01101        0.01179 

2. Quantiles for each variable:
 2.5%   25%   50%   75% 97.5% 
1.210 2.093 2.764 3.628 6.308
\end{verbatim}

\subsection{Proportional hazards models} \label{sec:PH}

The Cox proportional hazards model \citep{cox1972} expresses the hazard function $h(t)$ of the survival time of each individual of the target population as the product of a common baseline hazard function $h_{0}(t)$, which determines the shape of $h(t)$, and an exponential term which includes the relevant covariates ${\bm x}$ with regression coefficients $\bm \beta$ as follows:
\begin{equation}
h(t \mid h_{0}, {\bm \beta})= h_{0}(t)\exp\left\{{\bm x}^{\top}{\bm \beta}\right\}. \label{phmodel}
\end{equation}

The estimation of the regression coefficients in the Cox model under the frequentist approach can be obtained without specifying a model for the baseline hazard function by using partial likelihood methodology \citep{cox1972}. This is not the case of Bayesian analysis which in general needs to specify some model for the baseline hazard \citep{christensen2011}. Depending on the context of the study, baseline hazard misspecification can imply a loss of valuable model information that makes it impossible to fully report the estimation of the outcomes of interest, such as probabilities or survival curves for relevant covariate patterns \citep{royston2011}. This is specially important in survival studies where $h_{0}(t)$ represents the natural course of a disease or an infection, or even the control group when comparing several treatments.

Baseline hazard functions can be defined through parametric or semi-parametric approaches. Parametric models give restricted shapes which do not allow the presence of irregular behaviour. Semi-parametric choices result in more flexible baseline hazard shapes that allow for multimodal patterns such as piecewise constant functions or spline functions. Theoretical and methodological aspects of more flexible approaches to define the baseline hazard function can be found in several specific references such as \cite{ibrahim2001}, \cite{lin2015}, \cite{mitra2015}, \cite{bogaerts2017}, and \cite{lazaro2020}.

\subsubsection{Model specification}

The proportional hazards (PH) model implemented here is also illustrated with the \textit{larynx} dataset (see Section~\ref{sec:AFT}). Survival time for each individual is modelled by means of the hazard function: 
\begin{equation}
h(t \mid h_{0}, {\bm \beta}) = h_{0}(t) \exp\left\{ \beta_{2}\mbox{I}_{(\texttt{stage}=2)} + \beta_{3}\mbox{I}_{(\texttt{stage}=3)} + \beta_{4}\mbox{I}_{(\texttt{stage}=4)} + \beta_{5}\texttt{age} +\beta_{6}\texttt{diagyr}\right\}, \label{eq:larynx2}
\end{equation}

\noindent with baseline hazard function defined as a mixture of piecewise constant functions, $h_{0}(t \mid {\bm \lambda})= \sum_{k=1}^K\, \lambda_{k} \,I_{(a_{k-1},a_{k}]}(t)$, $t>0$, where ${\bm \lambda} = (\lambda_1, \ldots, \lambda_K)$ and $I_{(a_{k-1},a_{k}]}(t)$ is the indicator function defined as 1 when $t \in (a_{k-1},a_{k}]$ and 0 otherwise. We consider a total number of knots $K = 3$ and an equally-spaced partition of the time axis from $a_{0} = 0$ to $a_{4} = 10.70$ which corresponds to the longest survival time observed. Prior scenario is set under a non-informative independent framework with a N($0,0.001$) for $\beta$'s and an independent gamma distributions, Ga($0.01,0.01$), for each $\lambda$.

Piecewise constant baseline hazard functions can accommodate different shapes of the hazard depending on the particular characteristics of the partition of the time axis (See \cite{breslow1974}, \cite{murray2016}, and \cite{lee2016} for different proposals in the subject). Similarly, there is a wide range of approaches to define marginal prior distributions for $h_{0}$ parameters, from prior independence to prior correlation. Correlated scenarios account for shape restrictions and also avoid overfitting and strong irregularities in the estimation process \citep{lazaro2020}.

The likelihood function of this specific model is not implemented in \texttt{JAGS}, so the ``zeros trick'' approach using a Poisson distribution has been used to specify it indirectly \citep{ntzoufras2009, lunn2012}.

\subsubsection{Model implementation}

The variables \texttt{age}, \texttt{diagyr}, \texttt{stage}, and \texttt{X} are defined as in Section~\ref{sec:AFT}. In addition, as previously commented, piecewise constant is handled in \texttt{JAGS} considering the ``zeros trick'' \citep{ntzoufras2009, lunn2012}. To execute it, it is necessary to reformat the individual times (both observed and right-censored) to build an auxiliary variable (\texttt{int.obs}) that identifies the interval in which the $i$th individual experiences the event of interest. So, the following code is dedicated to define a time axis partition (i.e., intervals) and the \texttt{int.obs} variable:
\begin{verbatim}
R> # Time axis partition
R> K <- 3 # number of intervals
R> a <- seq(0, max(larynx$time) + 0.001, length.out = K + 1)
R> # int.obs: vector that tells us at which interval each observation is
R> int.obs <- matrix(data = NA, nrow = nrow(larynx), ncol = length(a) - 1)
R> d <- matrix(data = NA, nrow = nrow(larynx), ncol = length(a) - 1)
R> for(i in 1:nrow(larynx)){
+    for(k in 1:(length(a) - 1)){
+      d[i, k] <- ifelse(time[i] - a[k] > 0, 1, 0) * ifelse(a[k + 1] - time[i] > 0, 1, 0)
+      int.obs[i, k] <- d[i, k] * k
+    }
+  }
R> int.obs <- rowSums(int.obs)
\end{verbatim}

Listing~\ref{lst:ph} shows a generic implementation of a PH piecewise constant model in \texttt{BUGS} syntax using \textit{larynx} data.
\begin{figure}[htb!] \small
\begin{lstlisting}[caption={PH model in \texttt{BUGS} syntax (file named as \textbf{PH.txt}).}, label={lst:ph}]
model{
  for(i in 1:n){
    for(k in 1:int.obs[i]){
      cond[i,k] <- step(time[i] - a[k+1])
      HH[i,k] <- cond[i,k] * (a[k+1]-a[k]) * lambda[k] +
                 (1-cond[i,k]) * (time[i]-a[k]) * lambda[k]
    }
    # Cumulative hazard function
    H[i] <- sum(HH[i,1:int.obs[i]])
  }
  for(i in 1:n){
    # Linear predictor
    elinpred[i] <- exp(inprod(beta[],X[i,]))
    # Log-hazard function
    logHaz[i] <- log(lambda[int.obs[i]] * elinpred[i])
    # Log-survival function
    logSurv[i] <- -H[i] * elinpred[i]

    # Definition of the log-likelihood using zeros trick
    phi[i] <- 100000 - delta[i] * logHaz[i] - logSurv[i]
    zeros[i] ~ dpois(phi[i])
  }

  # Prior distributions
  for(l in 1:Nbetas){ beta[l] ~ dnorm(0,0.001) }
  for(k in 1:m){ lambda[k] ~ dgamma(0.01,0.01) }
}
\end{lstlisting}
\end{figure}

Once the variables have been defined, a list with all the elements required in the model is created:
\begin{verbatim}
R> d.jags <- list(n = nrow(larynx), m = length(a) - 1, delta = larynx$delta,
+    time = larynx$time, X = X[,-1], a = a, int.obs = int.obs, Nbetas = ncol(X) - 1,
+    zeros = rep(0, nrow(larynx)))
\end{verbatim}

The initial values for each PH model parameter are passed to \texttt{JAGS} using a function that returns a list of random values:
\begin{verbatim}
R> i.jags <- function(){ list(beta = rnorm(ncol(X) - 1), lambda = runif(3, 0.1)) }
\end{verbatim}

The vector of monitored/saved parameters is:
\begin{verbatim}
R> p.jags <- c("beta", "lambda")
\end{verbatim}

Next, the \texttt{JAGS} model is compiled:
\begin{verbatim}
R> library("rjags")
R> m2 <- jags.model(data = d.jags, file = "PH.txt", inits = i.jags, n.chains = 3)
\end{verbatim}

We now run the model for 1000 burn-in simulations:
\begin{verbatim}
R> update(m2, 1000)
\end{verbatim}

Finally, the model is run for 50000 additional simulations to keep one in 10 so that a proper thinning is done:
\begin{verbatim}
R> res <- coda.samples(m2, variable.names = p.jags, n.iter = 50000, n.thin = 10)
\end{verbatim}

Similarly to the first example (Section~\ref{sec:AFT}), numerical and graphical summaries of the model parameters can be obtained using the \texttt{summary} and \texttt{densplot} functions, respectively. Gelman and Rubin's convergence diagnostic can be calculated with the \texttt{gelman.diag} function, and the \texttt{traceplot} function provides a visual way to inspect sampling behaviour and assesses mixing across chains and convergence.

Next, simulations from the three Markov chains are merged together for inference:
\begin{verbatim}
R> result <- as.mcmc(do.call(rbind, res))
\end{verbatim}

The posterior samples of each parameter are obtained by:
\begin{verbatim}
R> beta2 <- result[,1]; beta3 <- result[,2]; beta4 <- result[,3]; beta5 <- result[,4]
R> beta6 <- result[,5]; lambda1 <- result[,6]; lambda2 <- result[,7]; lambda3 <- result[,8]
\end{verbatim}

Table~\ref{table:PH} shows posterior summaries for the PH model parameters using \textit{larynx} data.
\begin{table}[htb!] \centering
\caption{Posterior summaries for the PH model parameters. \label{table:PH}}
\begin{threeparttable}
\begin{tabular}{lcccccc}
\hline
Parameter &  Mean & SD & $2.5\%$ & $50\%$ & $97.5\%$ & $P(\cdot > 0 \mid \mbox{data})$ \\
\hline
$\beta_{2}$ (\texttt{stage}=2)  & 0.152 & 0.474 & -0.811 & 0.163 & 1.050 & 0.633 \\
$\beta_{3}$ (\texttt{stage}=3)  & 0.672 & 0.363 & -0.037 & 0.672 & 1.388 & 0.969 \\
$\beta_{4}$ (\texttt{stage}=4)  & 1.804 & 0.443 &  0.924 & 1.809 & 2.659 & 1.000 \\
$\beta_{5}$ (\texttt{age})      & 0.215 & 0.155 & -0.086 & 0.214 & 0.523 & 0.918 \\
$\beta_{6}$ (\texttt{diagyr})   &-0.042 & 0.167 & -0.368 &-0.042 & 0.288 & 0.400 \\
$\lambda_{1}$                   & 0.069 & 0.021 &  0.035 & 0.066 & 0.116 & 1.000 \\
$\lambda_{2}$                   & 0.104 & 0.035 &  0.048 & 0.100 & 0.185 & 1.000 \\
$\lambda_{3}$                   & 0.079 & 0.064 &  0.008 & 0.062 & 0.246 & 1.000 \\
\hline
\end{tabular}
\end{threeparttable}
\end{table}

The last column of Table~\ref{table:PH} contains the posterior probability that the corresponding parameter is positive. A probability equal to $0.5$ indicates that a positive value of the parameter is equally likely than a negative one. Once we have the posterior sample of each parameter stored, the calculation of this posterior probability, for example for $\beta_{2}$, is given by \texttt{mean(beta2>0)}.

A relevant quantity for PH models is the \textit{hazard ratio} (HR), also called \textit{relative risk}, between two individuals with covariate vectors ${\bm x}_{1}$ and ${\bm x}_{2}$. This measure is defined as: $$\mbox{HR}({\bm x}_{1},{\bm x}_{2},h_{0},{\bm \beta}) = \frac{h(t \mid {\bm x}_{1},h_{0},{\bm \beta})}{h(t \mid {\bm x}_{2},h_{0},{\bm \beta})} = \exp\left\{({\bm x}_{1} - {\bm x}_{2})^{\top}{\bm \beta}\right\},$$

\noindent and it is time independent. As an illustration, we can easily summarise the posterior distribution of the HR between two men of the same \texttt{age} and \texttt{diagyr} (year of diagnosis) but one in \texttt{stage}=3 and the other in \texttt{stage}=4:
\begin{verbatim}
R> HR.s3_s4 <- exp(beta3 - beta4)
R> summary(HR.s3_s4)
Iterations = 1:15000
Thinning interval = 1 
Number of chains = 1 
Sample size per chain = 15000 

1. Empirical mean and standard deviation for each variable,
   plus standard error of the mean:
          Mean             SD       Naive SE Time-series SE 
      0.354210       0.163810       0.001338       0.001338 

2. Quantiles for each variable:
  2.5%    25%    50%    75%  97.5% 
0.1384 0.2404 0.3217 0.4297 0.7667
\end{verbatim}

\section{Mixture cure models} \label{sec:Cure}

\subsection{Cure models} \label{sec:cure}

Cure models deal with target populations in which a part of the individuals cannot experience the event of interest. This type of models has widely been matured as a consequence of the discovery and development of new treatments against cancer. The rationale of considering a cure subpopulation comes from the idea that a successful treatment removes totally the original tumor and the individual cannot experience any recurrence of the disease. These models allow to estimate the probability of cure, a key and valuable outcome in cancer research.

Mixture cure models are the most popular cure models \citep{berkson1952}. They consider the target population as a mixture of susceptible and non-susceptible individuals for the event of interest. Let $Z$ be a cure random variable defined as $Z = 0$ for susceptible and $Z = 1$ for cured or immune individuals. Cure and non cure probabilities are $P(Z = 1) = \eta$ and $P(Z = 0) = 1 - \eta$, respectively. The survival function for each individual in the cured and uncured subpopulation, $S_{c}(t)$ and $S_{u}(t)$, $t > 0$, respectively, is
\begin{equation}
S_{u}(t) = P(T > t \mid Z = 0), \;\;\;\;\;\; S_{c}(t) = P(T > t \mid Z = 1),
\end{equation}
\noindent and the general survival function for $T$ can be expressed as $S(t) = P(T > t) = \eta + (1 - \eta) S_{u}(t)$. It is important to point out that $S_{u}(t)$ is a proper survival function but $S(t)$ is not. It goes to $\eta$ and not to zero when $t$ goes to infinity.

The effect of a baseline covariate vector ${\bm x}$ on the cure fraction $\eta$ for each individual is typically modelled by means of a logistic link function, logit$(\eta)$, but the probit link or the complementary log-log link could also be used. Covariates for modelling $T$ in the uncured subpopulation are usually considered via Cox models. Cure fraction model is usually known as the \textit{incidence model} and the survival model (i.e., time-to-event $T$ in the uncured subpopulation) as the \textit{latency model} \cite{klein2013}.

\subsection{\textit{bmt} dataset}

We consider a bone marrow transplant dataset, referred to as \textit{bmt}. It is available from the \texttt{smcure} package \citep{smcure}:
\begin{verbatim}
R> library("smcure")
R> data("bmt")
R> str(bmt)
'data.frame':	91 obs. of  3 variables:
 $ Time  : num  11 14 23 31 32 35 51 59 62 78 ...
 $ Status: num  1 1 1 1 1 1 1 1 1 1 ...
 $ TRT   : num  0 0 0 0 0 0 0 0 0 0 ...
\end{verbatim}

This dataset refers to a bone marrow transplant study for the refractory acute lymphoblastic leukemia patients, in which 91 patients were divided into two treatment groups \citep{kersey1987}. The following variables were observed for each patient:

\begin{itemize}
	\item \texttt{Time}: time to death (in days).
	\item \texttt{Status}: censoring indicator (1: if patient is uncensored; 0: otherwise).
	\item \texttt{TRT}: treatment group indicator (0: allogeneic; 1: autologous).
\end{itemize}

\subsection{Model specification}

The (cure subpopulation) incidence model for each individual is expressed in terms of a logistic regression:
\begin{equation}
 \mbox{logit}\big[\eta(\beta_{C1},\beta_{C2})\big] = \beta_{C1} + \beta_{C2}\texttt{TRT}, \label{eq:bmt1}
\end{equation}

\noindent where $\beta_{C1}$ represents an intercept and $\beta_{C2}$ is the regression coefficient for the \texttt{TRT} covariate.

Survival time for each individual in the uncured subpopulation is modelled from a proportional hazards specification:
\begin{equation}
h_{u}(t \mid h_{0},\beta_{U}) = h_{0}(t)\exp\left\{\beta_{U}\texttt{TRT}\right\}, \label{eq:bmt2}
\end{equation}

\noindent with $h_{0}(t)=\lambda \, \alpha \, t^{\alpha-1}$ specified as a Weibull baseline hazard function, where $\alpha$ and $\lambda$ are the shape and scale parameters, respectively; and $\beta_{U}$ is the regression coefficient for the \texttt{TRT} covariate.

We assume prior independence and specify prior marginal distributions based on non-informative distributions commonly employed in the literature. The $\beta$'s follow a N($0,0.001$), while $\lambda$ and $\alpha$ follow a Gamma($0.01,0.01$) and a Un($0,10$), respectively.

The likelihood function for mixture cure models is not implemented in \texttt{JAGS}, so the ``zeros trick'' approach using a Poisson distribution has also been used to specify it indirectly \citep{ntzoufras2009, lunn2012}.

\subsection{Model implementation}

We have created two design matrices, one for model (\ref{eq:bmt1}) and another for model (\ref{eq:bmt2}), with the \texttt{TRT} covariate:
\begin{verbatim}
R> XC <- model.matrix(~ TRT, data = bmt) # Reference = allogeneic
R> XU <- model.matrix(~ TRT, data = bmt)
R> XU <- matrix(XU[,-1], ncol = 1) # Remove intercept
\end{verbatim}

Listing~\ref{lst:cure} shows a generic implementation of a mixture cure model in \texttt{BUGS} syntax using \textit{bmt} data.

\begin{figure}[htb!] \small
\begin{lstlisting}[caption={Mixture cure model in \texttt{BUGS} syntax (file named as \textbf{Cure.txt}).}, label={lst:cure}]
model{
  for(i in 1:n){
    # Logistic regression model (cured subpopulation)
    logit(eta[i]) <- inprod(betaC[],XC[i,])

    # PH model (uncured subpopulation)
    # Weibull baseline
    base[i] <- lambda * alpha * pow(t[i],alpha-1)
    elinpred[i] <- exp(inprod(betaU[],XU[i,]))
    # Log-hazard function
    logHaz[i] <- log(base[i] * elinpred[i])
    # Log-survival function
    logSurv[i] <- -lambda * pow(t[i],alpha) * elinpred[i]

    # Definition of the log-likelihood using zeros trick
    logLike[i] <- delta[i] * (log(1-eta[i]) + logHaz[i] + logSurv[i]) +
               (1-delta[i]) * log(eta[i] + (1-eta[i]) * exp(logSurv[i]))
    phi[i] <- 100000 - logLike[i]
    zeros[i] ~ dpois(phi[i])
  }

  # Prior distributions
  for(l in 1:NbetasC){ betaC[l] ~ dnorm(0,0.001) }
  for(l in 1:NbetasU){ betaU[l] ~ dnorm(0,0.001) }
  lambda ~ dgamma(0.01,0.01)
  alpha ~ dunif(0,10)
}
\end{lstlisting}
\end{figure}

Once the variables have been defined, a list with all the elements required in the model is created:
\begin{verbatim}
R> d.jags <- list(n = nrow(bmt), t = bmt$Time, XC = XC, XU = XU,
+    delta = bmt$Status, zeros = rep(0, nrow(bmt)), NbetasC = ncol(XC), NbetasU = ncol(XU))
\end{verbatim}

The initial values for each mixture cure model parameter are passed to \texttt{JAGS} using a function that returns a list of random values:
\begin{verbatim}
R> i.jags <- function(){
+    list(betaC = rnorm(ncol(XC)), betaU = rnorm(ncol(XU)), lambda = runif(1), alpha = runif(1))
+  }
\end{verbatim}

The vector of monitored/saved parameters is:
\begin{verbatim}
R> p.jags <- c("betaC", "betaU", "alpha", "lambda")
\end{verbatim}

Next, the \texttt{JAGS} model is compiled:
\begin{verbatim}
R> library("rjags")
R> m3 <- jags.model(data = d.jags, file = "Cure.txt", inits = i.jags, n.chains = 3)
\end{verbatim}

We now run the model for 10000 burn-in simulations:
\begin{verbatim}
R> update(m3, 10000)
\end{verbatim}

Finally, the model is run for 100000 additional simulations to keep one in 100 so that a proper thinning is done:
\begin{verbatim}
R> res <- coda.samples(m3, variable.names = p.jags, n.iter = 100000, n.thin = 100)
\end{verbatim}

Similarly to the first example (Section~\ref{sec:AFT}), numerical and graphical summaries of the model parameters can be obtained using the \texttt{summary} and \texttt{densplot} functions, respectively. Gelman and Rubin's convergence diagnostic can be calculated with the \texttt{gelman.diag} function, and the \texttt{traceplot} function provides a visual way to inspect sampling behaviour and assesses mixing across chains and convergence.

Next, simulations from the three Markov chains are merged together for inference:
\begin{verbatim}
R> result <- as.mcmc(do.call(rbind, res))
\end{verbatim}

The posterior samples of each parameter are obtained by:
\begin{verbatim}
R> alpha <- result[,1]; betaC1 <- result[,2]; betaC2 <- result[,3]
R> betaU <- result[,4]; lambda <- result[,5]
\end{verbatim}

Table~\ref{table:cure} shows posterior summaries for the mixture cure model parameters using \textit{bmt} data.

\begin{table}[htb!] \centering
\caption{Posterior summaries for the mixture cure model parameters. \label{table:cure}}
\begin{threeparttable}
\begin{tabular}{lcccccc}
\hline
Parameter &  Mean & SD & $2.5\%$ & $50\%$ & $97.5\%$ & $P(\cdot > 0 \mid \mbox{data})$ \\
\hline
$\beta_{C1}$ (\texttt{intercept}) &  -1.015  & 0.349  &  -1.731  &  -1.002  & -0.365  & 0.001 \\
$\beta_{C2}$ (\texttt{TRT})       &  -0.419  & 0.519  &  -1.445  &  -0.417  &  0.591  & 0.208 \\
$\beta_{U}$ (\texttt{TRT})        &   0.762  & 0.269  &   0.239  &   0.760  &  1.294  & 0.998 \\
$\alpha$                          &   1.143  & 0.105  &   0.943  &   1.140  &  1.354  & 1.000 \\
$\lambda$                         &   0.002  & 0.001  &   0.000  &   0.002  &  0.006  & 1.000 \\
\hline
\end{tabular}
\end{threeparttable}
\end{table}

As we discussed earlier, the cure fraction ($\eta$) is a relevant quantity for mixture cure models. For allogeneic (\texttt{TRT}=0) or autologous (\texttt{TRT}=1) treated patients, it is modelled as: $$\eta(\texttt{TRT},\beta_{C1},\beta_{C2}) = \frac{\exp(\beta_{C1} + \beta_{C2}\texttt{TRT})}{1+\exp(\beta_{C1} + \beta_{C2}\texttt{TRT})}.$$

We can easily summarise the posterior distribution of the cure fraction for individuals in both groups:
\begin{verbatim}
R> CP.allo <- exp(betaC1) / (1 + exp(betaC1))
R> CP.auto <- exp(betaC1 + betaC2) / (1 + exp(betaC1 + betaC2))
R> summary(cbind(CP.allo, CP.auto))
    CP.allo           CP.auto       
 Min.   :0.01604   Min.   :0.02277  
 1st Qu.:0.22426   1st Qu.:0.15680  
 Median :0.26850   Median :0.19477  
 Mean   :0.27146   Mean   :0.19925  
 3rd Qu.:0.31515   3rd Qu.:0.23699  
 Max.   :0.65573   Max.   :0.55898
\end{verbatim}

\begin{figure}[bt]
\centering
\includegraphics[scale=0.6]{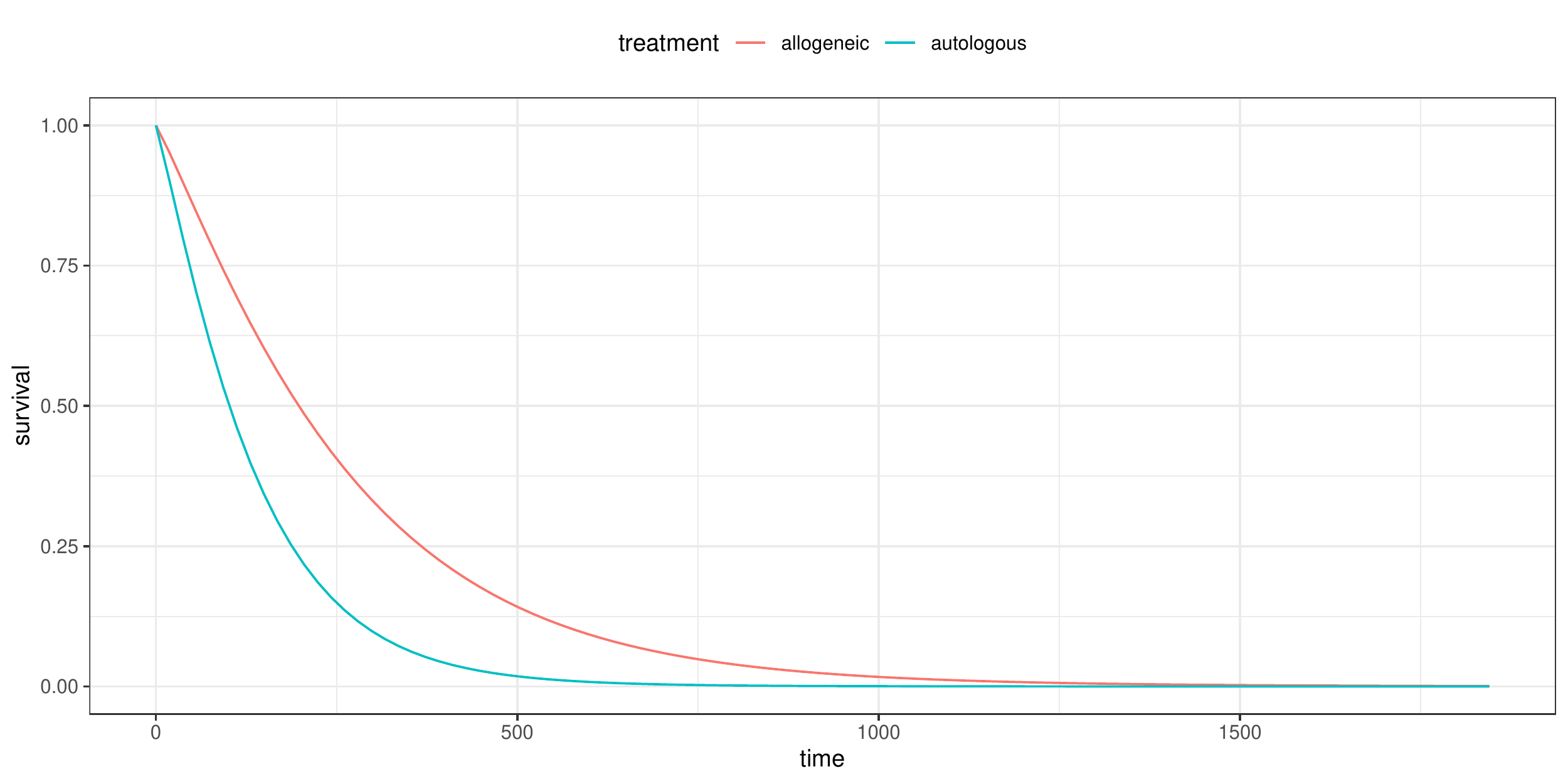}
\caption{\label{fig:cure}Posterior mean of the uncured survival function from the mixture cure model (\ref{eq:bmt2}).}
\end{figure}

The uncured survival curve based on posterior samples is another relevant information in this type of studies. So, from the posterior samples obtained above, we can summarise the posterior distribution of the mean value of the uncured survival curve for allogeneic and autologous treated patients in a grid of points as follows:
\begin{verbatim}
R> grid <- 100
R> time <- seq(0, bmt$Time, len = grid)
R> surv.allo <- surv.auto <- vector()
R> for(l in 1:grid){
+     surv.allo[l] <- mean(exp(-lambda * time[l]^alpha))
+     surv.auto[l] <- mean(exp(-lambda * exp(betaU) * time[l]^alpha))
+  }
\end{verbatim}

Figure~\ref{fig:cure} shows the difference between both curves using the code below:
\begin{verbatim}
R> library("ggplot2")
R> treat.col <- rep(0:1, each = grid)
R> treat.col[treat.col == 0] <- "allogeneic"
R> treat.col[treat.col == 1] <- "autologous"
R> df <- data.frame(time = rep(time, 2), survival = c(surv.allo, surv.auto),
+    treatment = treat.col)
R> ggplot(data = df, aes(x = time, y = survival, group = treatment, colour = treatment)) +
+    geom_line() + theme_bw() + theme(legend.position = "top")
\end{verbatim}

\section{Competing risks models} \label{sec:CR}

Competing risks occur when the survival process includes more than one cause of failure. In the case of different causes of death it is only possible to report the first event to occur \citep{putter2007}. There are different approaches for competing risk models: multivariate time to failure model, the cause-specific hazards model, the mixture model, the subdistribution model, and the full specified subdistribution model \citep{ge2012}. We will only consider here the cause-specific hazards model, possibly the most popular of them.

Let $T_{k}$ be the random variable that represents the time-to-event from cause $k$, for $k=1,\ldots,K$, where $K$ is the total number of different events. The only survival time observed $T=\min\{T_{1}, T_{2},\ldots,T_{K}\}$ usually corresponds to the earliest cause together with their indicator $\delta=k$ when the subsequent individual experiences the event due to cause $k$. The key concept in a competing risk model is the cause-specific hazard function for cause $k$, which assesses the hazard of experiencing the event $k$ in the presence of the rest of competing events, and it is defined by:
\begin{equation}
h_{k}(t) = \lim_{\Delta t \to 0} \frac{P(t \leq T < t+\Delta t, \delta=k \mid T \geq t)}{\Delta t}. \label{comprisksmodel}
\end{equation}

Inference for each $h_{k}(t)$ considers the observed failure times for cause $k$ as censored observations for the rest of events. Another relevant concept in the competing framework is the cumulative incidence function for cause $k$, defined as follows:
\begin{equation}
F_{k}(t)=P(T \leq t, \delta=k) = \int_{0}^{t}\, h_{k}(u)\,S(u)\,\mbox{d}u, \label{cif}
\end{equation}

\noindent where $S(t)$ is the overall survival function. It is typically expressed in terms of the different cause-specific hazard functions in accordance with
$$S(t)=P(T > t)=\exp\Big\{-\sum_{k=1}^{K}\, \int_{0}^{t} \, h_{k}(u)\,\mbox{d}u \Big\}.$$

The cumulative incidence function is not a proper cumulative distribution function, i.e., the probability that the subsequent individual fails from cause $k$ is $F_{k}(\infty)=P(\delta=k) \neq 1$. The sub-survival function for cause $k$, defined as $S_{k}(t)=P(T>t, \delta=k)$, is also not a proper survival function.

\subsection{\textit{okiss} dataset}

We consider a stem-cell transplanted patients dataset, referred to as \textit{okiss}. It is available from the \texttt{compeir} package \citep{compeir}:
\begin{verbatim}
R> library("compeir")
R> data("okiss")
R> str(okiss)
'data.frame':	1000 obs. of  4 variables:
 $ time  : 'times' num  21 6 14 12 11 18 10 8 17 10 ...
  ..- attr(*, "format")= chr "h:m:s"
 $ status: num  2 1 2 2 2 2 7 2 2 2 ...
 $ allo  : num  1 1 1 1 1 1 1 0 1 1 ...
 $ sex   : Factor w/ 2 levels "f","m": 1 1 1 2 2 2 2 1 2 2 ...
\end{verbatim}

This dataset provides information about a sub-sample of 1000 patients enrolled in the ONKO-KISS programme, which is part of the German National Reference Centre for Surveillance of Hospital-Acquired Infections \citep{dettenkofer2005}. These patients have been treated by peripheral blood stem-cell transplantation, which has become a successful therapy for severe hematologic diseases. After transplantation, patients are neutropenic, i.e., they have a low count of white blood cells, which are the cells that primarily avert infections \citep{beyersmann2007}. Occurrence of bloodstream infection during neutropenia is a severe complication. The following variables were observed for each patient:
\begin{itemize}
	\item \texttt{time}: time (in days) of neutropenia until first event.
	\item \texttt{status}: event status indicator (1: infection; 2: end of neutropenia; 7: death; 11: censored observation).
	\item \texttt{allo}: transplant type indicator (0: autologous; 1: allogeneic).
	\item \texttt{sex}: sex of each patient (m: if patient is male; f: if patient is female).
\end{itemize}

We have redefined the \texttt{status} variable using an auxiliary one (\texttt{delta}) in a matrix format:
\begin{verbatim}
R> delta <- matrix(c(as.integer(okiss$status == 1), as.integer(okiss$status == 2),
+    as.integer(okiss$status == 7)), ncol = 3)
R> head(delta)
     [,1] [,2] [,3]
[1,]    0    1    0
[2,]    1    0    0
[3,]    0    1    0
[4,]    0    1    0
[5,]    0    1    0
[6,]    0    1    0
\end{verbatim}

\noindent where the events 1 (infection), 2 (end of neutropenia) and 7 (death) are indicated with a value of 1 in column 1, 2 or 3, respectively, and a row with only 0's represents a censored observation.

\subsection{Model specification}

Cause-specific hazard functions for infection ($k=1$), end of neutropenia ($k=2$), and death ($k=3$) are modelled from a proportional hazard specification:
\begin{equation}
h_{k}(t \mid h_{0k}, {\bm \beta}_{k}) = h_{0k}(t)\exp\left\{\beta_{1k}\texttt{allo} + \beta_{2k}\texttt{sex} \right\}, \;\;\; k=1,2,3, \label{eq:okiss}
\end{equation}

\noindent with $h_{0k}(t)=\lambda_{k} \, \alpha_{k} \, t^{\alpha_{k}-1}$ for event $k$ specified as a Weibull baseline hazard function, where $\alpha_{k}$ and $\lambda_{k}$ are the shape and scale parameters, respectively; and ${\bm \beta}_{k}=(\beta_{1k},\beta_{2k})^{\top}$ are regression coefficients for the \texttt{allo} and \texttt{sex} covariates, respectively, for $k=1,2,3$. We assume prior independence and specify prior marginal based on non-informative distributions commonly employed in the literature. The $\beta$'s follow a N($0,0.001$), while $\lambda$'s and $\alpha$'s follow a Gamma($0.01,0.01$) and a Un($0,10$), respectively.

\subsection{Model implementation}

We have created a design matrix \texttt{X} with the covariates \texttt{allo} and \texttt{sex}:
\begin{verbatim}
R> X <- model.matrix(~ allo + sex, data = okiss) # Reference = female
R> X <- X[,-1] # Remove intercept
\end{verbatim}

Listing~\ref{lst:comprisks} shows a generic implementation of a competing risks model in \texttt{BUGS} syntax using \textit{okiss} data.

\begin{figure}[htb!] \small
\begin{lstlisting}[caption={Competing risks model in \texttt{BUGS} syntax (file named as \textbf{CR.txt}).}, label={lst:comprisks}]
model{
  for(i in 1:n){
    for(k in 1:Nrisks){
      # Weibull baseline
      base[i,k] <- lambda[k] * alpha[k] * pow(t[i],alpha[k]-1)
      elinpred[i,k] <- exp(inprod(beta[,k],X[i,]))
      # Log-hazard functions
      logHaz[i,k] <- log(base[i,k] * elinpred[i,k])
      # Log-survival functions
      logSurv[i,k] <- -lambda[k] * pow(t[i],alpha[k]) * elinpred[i,k]
    }

    # Definition of the log-likelihood using zeros trick
    phi[i] <- 100000 - inprod(delta[i,],logHaz[i,]) - sum(logSurv[i,])
    zeros[i] ~ dpois(phi[i])
  }

  # Prior distributions
  for(k in 1:Nrisks){
    for(l in 1:Nbetas){ beta[l,k] ~ dnorm(0,0.001) }
    lambda[k] ~ dgamma(0.01,0.01)
    alpha[k] ~ dunif(0,10)
  }
}
\end{lstlisting}
\end{figure}

Once the variables have been defined, a list with all the elements required in the model is created:
\begin{verbatim}
R> d.jags <- list(n = nrow(okiss), t = as.vector(okiss$time), X = X,
+    delta = delta, zeros = rep(0, nrow(okiss)), Nbetas = ncol(X), Nrisks = ncol(delta))
\end{verbatim}

The initial values for each competing risks model parameter are passed to \texttt{JAGS} using a function that returns a list of random values:
\begin{verbatim}
R> i.jags <- function(){
+    list(beta = matrix(rnorm(ncol(X) * ncol(delta)), ncol = ncol(delta)),
+      lambda = runif(ncol(delta)), alpha = runif(ncol(delta)))
+  }
\end{verbatim}

The vector of monitored/saved parameters is:
\begin{verbatim}
R> p.jags <- c("beta", "alpha", "lambda")
\end{verbatim}

Next, the \texttt{JAGS} model is compiled:
\begin{verbatim}
R> library("rjags")
R> m4 <- jags.model(data = d.jags, file = "CR.txt", inits = i.jags, n.chains = 3)
\end{verbatim}

We now run the model for 1000 burn-in simulations:
\begin{verbatim}
R> update(m4, 1000)
\end{verbatim}

Finally, the model is run for 10000 additional simulations to keep one in 10 so that a proper thinning is done:
\begin{verbatim}
R> res <- coda.samples(m4, variable.names = p.jags, n.iter = 10000, n.thin = 10)
\end{verbatim}

Similarly to the first example (Section~\ref{sec:AFT}), numerical and graphical summaries of the model parameters can be obtained using the \texttt{summary} and \texttt{densplot} functions, respectively. Gelman and Rubin's convergence diagnostic can be calculated with the \texttt{gelman.diag} function, and the \texttt{traceplot} function provides a visual way to inspect sampling behaviour and assesses mixing across chains and convergence.

Next, simulations from the three Markov chains are merged together for inference:
\begin{verbatim}
R> result <- as.mcmc(do.call(rbind, res))
\end{verbatim}

The posterior samples of each parameter are obtained by:
\begin{verbatim}
R> alpha1 <- result[,1]; alpha2 <- result[,2]; alpha3 <- result[,3]; beta11 <- result[,4]
R> beta21 <- result[,5]; beta12 <- result[,6]; beta22 <- result[,7]; beta13 <- result[,8]
R> beta23 <- result[,9]; lambda1 <- result[,10]; lambda2 <- result[,11]; lambda3 <- result[,12]
\end{verbatim}

Table~\ref{table:CR} shows posterior summaries for the competing risks model parameters using \textit{okiss} data.
\begin{table}[htb!] \centering
\caption{Posterior summaries for the competing risks model parameters. \label{table:CR}}
\begin{threeparttable}
\begin{tabular}{lcccccc}
\hline
Parameter &  Mean & SD & $2.5\%$ & $50\%$ & $97.5\%$ & $P(\cdot > 0 \mid \mbox{data})$ \\
\hline
\multicolumn{7}{c}{Infection ($h_{1}$)} \\
\hline
$\beta_{11}$ (\texttt{allo})   & -0.523  & 0.151  & -0.817  & -0.523  & -0.225  & 0.000 \\
$\beta_{21}$ (\texttt{sex})    &  0.158  & 0.148  & -0.130  &  0.156  &  0.451  & 0.857 \\
$\lambda_{1}$                  &  0.014  & 0.003  &  0.009  &  0.014  &  0.022  & 1.000 \\
$\alpha_{1}$                   &  1.137  & 0.070  &  1.003  &  1.136  &  1.278  & 1.000 \\
\hline
\multicolumn{7}{c}{End of neutropenia ($h_{2}$)} \\
\hline
$\beta_{12}$ (\texttt{allo})   & -1.193  & 0.075  & -1.340  & -1.193  & -1.046  & 0.000 \\
$\beta_{22}$ (\texttt{sex})    & -0.102  & 0.073  & -0.244  & -0.103  &  0.042  & 0.081 \\
$\lambda_{2}$                  &  0.008  & 0.001  &  0.006  &  0.008  &  0.010  & 1.000 \\
$\alpha_{2}$                   &  2.033  & 0.045  &  1.947  &  2.034  &  2.120  & 1.000 \\
\hline
\multicolumn{7}{c}{Death ($h_{3}$)} \\
\hline
$\beta_{13}$ (\texttt{allo})   & -0.617  & 0.747  & -2.001  & -0.650  &  0.945  & 0.193 \\
$\beta_{23}$ (\texttt{sex})    &  0.446  & 0.730  & -0.859  &  0.408  &  1.986  & 0.718 \\
$\lambda_{3}$                  &  0.000  & 0.000  &  0.000  &  0.000  &  0.000  & 1.000 \\
$\alpha_{3}$                   &  2.628  & 0.418  &  1.813  &  2.623  &  3.466  & 1.000 \\
\hline
\end{tabular}
\end{threeparttable}
\end{table}

As discussed in Section~\ref{sec:CR}, the cumulative incidence function $F_{k}(t)$ in (~\ref{cif}) is the most appropriate way to analyse the evolution of each cause $k$ over time. For our proportional hazard specification~(\ref{eq:okiss}), it is given by:
\begin{equation}
F_{k}(t) = \int_{0}^{t} \, h_{k}(t \mid h_{0k}, {\bm \beta}_{k})\,\exp\Big\{-\sum_{l=1}^{3}\int_{0}^{u}\, h_{l}(v \mid h_{0l}, {\bm \beta}_{l})\,\mbox{d}v \Big\} \,\mbox{d}u, \label{weibullcif}
\end{equation}

\noindent where $h_{k}(t \mid h_{0k}, {\bm \beta}_{k})$ is defined in (\ref{eq:okiss}).

The integral in (\ref{weibullcif}) has no closed form, so some approximate method of integration is required. To do this, we first have created a function \texttt{fk} which describes the integrand of (\ref{weibullcif}):
\begin{verbatim}
R> fk <- function(u.vect, lambda, alpha, beta, x, k){
+  res <- sapply(u.vect, function(u){
+    # Cause-specific hazard
+    hk <- lambda[k] * alpha[k] * (u^(alpha[k] - 1)) * exp(sum(unlist(beta[,k]) * x))
+    # Cumulative cause-specific hazard
+    Hk <- lambda * (rep(u, length(lambda))^alpha) * exp((t(beta) %*% matrix(x, ncol = 1))[,1])
+    # Cause-specific hazard x Overall survival
+    aux <- hk * exp(-sum(Hk))
+    return(aux)
+  })
+  return(res) }
\end{verbatim}

Next, we have created a function \texttt{cif} that computes $F_{k}(t)$ in (\ref{weibullcif}) by integrating out the \texttt{fk} function using the \texttt{integral} function available from the \texttt{pracma} package \citep{pracma}.
\begin{verbatim}
R> library("pracma")
R> cif <- function(tt, lambda, alpha, beta, x, k){
+    return(integral(fk, xmin = 0, xmax = tt, method = "Simpson", lambda = lambda, alpha = alpha,
+      beta = beta, x = x, k = k)) }
\end{verbatim}

Finally, we have constructed a function \texttt{mcmc$\_$cif} that takes the output from \texttt{JAGS} (variable \texttt{obj}) and computes $F_{k}(t)$ in (\ref{weibullcif}) for a vector of times (variable \texttt{t.pred}) using covariates \texttt{x}. Note that \texttt{mcmc$\_$cif} is based on the \texttt{mclapply} function, available from the \texttt{parallel} package to speed computations up \citep{R}.

\begin{verbatim}
R> library("parallel")
R> options(mc.cores = detectCores())
R> mcmc_cif <- function(obj, t.pred, x){
+    var.names <- names(obj)
+    # Indices of beta's, alpha's, and lambda's
+    b.idx <- which(substr(var.names, 1, 4) == "beta")
+    a.idx <- which(substr(var.names, 1, 5) == "alpha")
+    l.idx <- which(substr(var.names, 1, 6) == "lambda")
+    # Number of causes and number of covariates
+    K <- length(a.idx)
+    n.b <- length(b.idx) / K
+    # Sub-sample to speed up computations
+    samples.idx <- sample(1:nrow(obj), 200)
+
+    res <- lapply(1:K, function(k){
+      sapply(t.pred, function(tt){
+        aux <- mclapply(samples.idx, function(i){
+          cif(tt, alpha = unlist(obj[i, a.idx]), lambda = unlist(obj[i, l.idx]),
+             beta = matrix(unlist(c(res[i, b.idx])), nrow = n.b), x = x, k = k)
+        })
+        return(mean(unlist(aux)))
+      })
+    })
+    return(res)
+  }
\end{verbatim}

Hence, we redefine the MCMC output as a \texttt{data.frame} and set a vector of times to evaluate the cumulative incidence function. In this example, we are interested in this function  when both covariates are 1 (i.e., allogeneic transplant and male).
\begin{verbatim}
R> res <- as.data.frame(result)
R> t.pred <- seq(0, 100, by = 2.5)
R> cum_inc <- mcmc_cif(res, t.pred, c(1, 1))
\end{verbatim}

\begin{figure}[bt]
\centering
\includegraphics[scale=0.6]{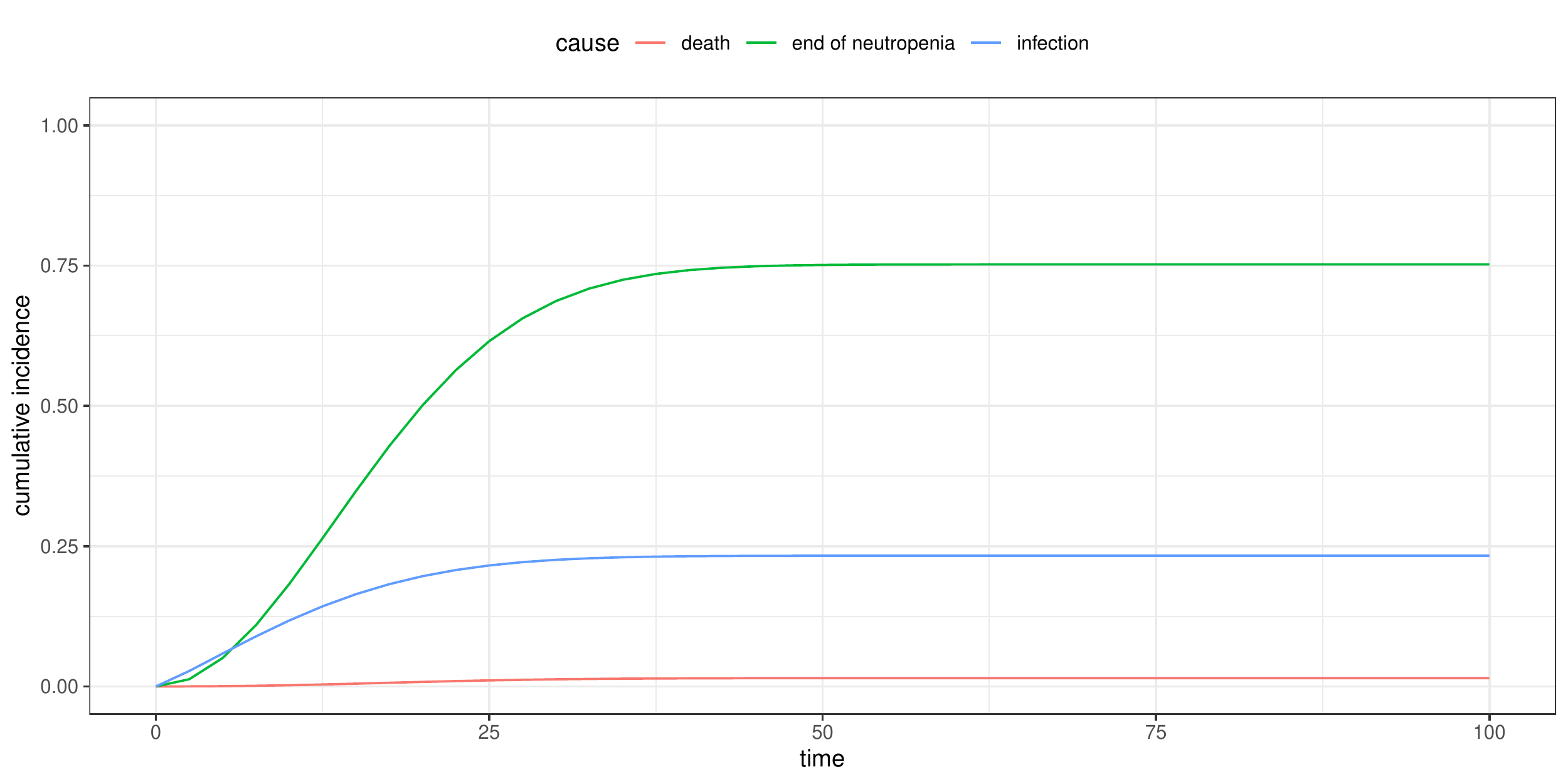}
\caption{\label{fig:cuminc}Posterior mean of the cumulative incidence function for \texttt{allo}=1 (allogeneic transplant) and \texttt{sex}=1 (male) from the competing risks model (\ref{eq:okiss}).}
\end{figure}

Figure~\ref{fig:cuminc}, generated with the code below, shows the posterior mean of the cumulative incidence function for a man with an allogeneic transplant for the three types of events considered in the \textit{okiss} data.
\begin{verbatim}
R> library("ggplot2")
R> df <- data.frame(cif = unlist(cum_inc), time = t.pred,
+    cause = rep(c("infection", "end of neutropenia", "death"), each = length(t.pred)))
R> ggplot(data = df, aes(x = time, y = cif, group = cause)) + geom_line(aes(color = cause)) +
+    ylab("cumulative incidence") + ylim(c(0,1)) + theme_bw() + theme(legend.position = "top")
\end{verbatim}

\section{Multi-state models} \label{sec:MS}

Multi-state models are a class of stochastic processes which account for event history data, with individuals who may experience different events in time. Relevant data are the events and their subsequent survival times. Multi-state models allow for different structures depending on the number and relationships between the states \citep{andersen2002}. Outputs of interest are the usual in survival analysis (sometimes with a specific vocabulary, e.g., the hazard function which is now called \textit{transition intensity}) to which transition probabilities are added.

We concentrate on the illness-death model (also known as \textit{disability model}) which is a particular multi-state model. This is a relevant model in irreversible diseases where a significant illness' progression increases the risk of a terminal event \citep{han2014, armero2016b}. The underlying stochastic process $\{Z(t), t \geq 0\}$ describes the state of an individual at time $t$, where $t$ is time from entry into the initial state (state 1). The probabilistic behaviour of the process is determined by the subsequent hazard functions from which transitions probabilities between states, defined as $p_{jk}(s,t \mid \boldsymbol \theta)= P(Z(t)=k \mid Z(s)=j, \boldsymbol \theta)$, can be derived, where $s \leq t$, $j$ and $k$ are states, $\sum_{j=k}^{3}\,p_{jk}(s,t \mid \boldsymbol \theta)=1$, for $k=1,2,3$, and $\boldsymbol \theta$ is the vector of model parameters.

Each transition has its respective hazard function, for example, $h_{12}(t \mid \boldsymbol \theta)$ is associated with time $T_{12}$ from state 1 to state 2 (1 $\rightarrow$ 2), while $h_{13}(t \mid \boldsymbol \theta)$ and $h_{23}(t \mid \boldsymbol \theta)$ represent the hazard functions for the time $T_{13}$ (1 $\rightarrow$ 3) and $T_{23}$ (2 $\rightarrow$ 3), respectively. If $h_{23}(t \mid \boldsymbol \theta)$ does not depend of the time in which transition 1 $\rightarrow$ 2 occurs, the process will be known as {\it Markovian}. Otherwise, it is called {\it semi-Markovian} and transition probabilities 2 $\rightarrow$ 3 will depend on the particular value of $T_{12}$.

Transition probabilities between states and hazard functions for the semi-Markovian specification are connected as follows \citep{andersen2008}:
\begin{eqnarray}
  p_{11}(s, t \mid \boldsymbol \theta) &=& \exp\Bigg\{-\int_{s}^{t}\left[h_{12}(u \mid \boldsymbol \theta)+h_{13}(u \mid \boldsymbol \theta)\right]\mbox{d}u \Bigg\}, \label{eq:tp11}\\
  p_{22}(s, t \mid \boldsymbol \theta, t_{12}) &=& \exp\Bigg\{-\int_{s}^{t}\,h_{23}(u-t_{12} \mid \boldsymbol \theta, t_{12}) \;\mbox{d}u \Bigg\}, \;\;\; t_{12}<s, \label{eq:tp22} \\
  p_{12}(s, t \mid \boldsymbol \theta) &=& \int_{s}^{t} \,p_{11}(s,u \mid \boldsymbol \theta) \,h_{12}(u \mid \boldsymbol \theta)\,p_{22}(u,t \mid \boldsymbol \theta, u) \;\mbox{d}u, \label{eq:tp12} \\
	p_{13}(s, t \mid \boldsymbol \theta) &=& 1-p_{11}(s, t \mid \boldsymbol \theta)-p_{12}(s, t \mid \boldsymbol \theta), \label{eq:tp13} \\
	p_{23}(s, t \mid \boldsymbol \theta) &=& 1-p_{22}(s, t \mid \boldsymbol \theta), \label{eq:tp23} \\
	p_{33}(s, t \mid \boldsymbol \theta) &=& 1, \label{eq:tp33}
\end{eqnarray}

\noindent where the marginal probability $p_{22}(s, t \mid \boldsymbol \theta)$ is obtained by integrating out $p_{22}(s, t \mid \boldsymbol \theta, t_{12})$ with regard to the density of $T_{12}$ from 0 to $s$.

The standard specification of an illness-death model is through the hazard function of the relevant survival times, generally modelled by means of Cox models. Consequently, the specification of prior distributions for $\boldsymbol \theta$ must be addressed to them in Section~\ref{sec:PH}.

\subsection{\textit{heart2} dataset}

We consider a heart transplant dataset, referred to as \textit{heart2}. It is available from the \texttt{p3state.msm} package \citep{p3state.msm}:
\begin{verbatim}
R> library("p3state.msm")
R> data("heart2")
R> str(heart2)
'data.frame':	103 obs. of  8 variables:
 $ times1 : num  50 6 1 36 18 3 51 40 85 12 ...
 $ delta  : int  0 0 1 1 0 0 1 0 0 1 ...
 $ times2 : num  0 0 15 3 0 0 624 0 0 46 ...
 $ time   : int  50 6 16 39 18 3 675 40 85 58 ...
 $ status : int  1 1 1 1 1 1 1 1 1 1 ...
 $ age    : num  -17.16 3.84 6.3 -7.74 -27.21 ...
 $ year   : num  0.123 0.255 0.266 0.49 0.608 ...
 $ surgery: int  0 0 0 0 0 0 0 0 0 0 ...
\end{verbatim}

This dataset provides information about a sample of 103 patients of the Stanford Heart Transplant Program \citep{crowley1977}. The patients are initially on the waiting list (state 1) and can either be transplanted (state 2, non-terminal event) and then die (state 3, terminal event), or just one or none of them because they continue to be on the waiting list. The following variables were observed for each patient:
\begin{itemize}
	\item \texttt{times1}: time of transplant/censoring time (state 2).
	\item \texttt{delta}: transplant indicator (1: yes; 0: no).
	\item \texttt{times2}: time to death since the transplant/censoring time (state 3).
	\item \texttt{time}: \texttt{times1} + \texttt{times2}.
	\item \texttt{status}: censoring indicator (1: dead; 0: alive).
	\item \texttt{age}: age - 48 years.
	\item \texttt{year}: year of acceptance (in years after 1 Nov 1967).
	\item \texttt{surgery}: prior bypass surgery (1: yes; 0: no).		
\end{itemize}

The patients had the following characteristics: 4 were censored for both events (transplant and death), 24 moved from state 1 (waiting list) to state 2 (transplant) and survived, 30 moved from state 1 to state 3 (death) without going through state 2, and 45 moved from state 1 to state 2 and then to state 3.

We have redefined \texttt{delta} and \texttt{status} variables using an auxiliary one (\texttt{event}) in a matrix format:
\begin{verbatim}
R> event <- matrix(c(heart2$delta, heart2$status * (1 - heart2$delta),
+    heart2$delta * heart2$status), ncol = 3)
R> head(event)
     [,1] [,2] [,3]
[1,]    0    1    0
[2,]    0    1    0
[3,]    1    0    1
[4,]    1    0    1
[5,]    0    1    0
[6,]    0    1    0
\end{verbatim}

\noindent where each row represents the transitions of a patient, in which a 1 in columns 1, 2 and 3 indicates transition 1 $\rightarrow$ 2, 1 $\rightarrow$ 3, and 2 $\rightarrow$ 3, respectively, and a row with only 0's represents a censored observation.

\subsection{Model specification}

Hazard functions for survival times $T_{12}$, $T_{13}$ and $T_{23}$ are modelled using a proportional hazard specification:
\begin{eqnarray}
	h_{12}(t \mid h_{01}, {\bm\beta}_{1}) &=& h_{01}(t)\exp\left\{\beta_{11}\texttt{age}+\beta_{21}\texttt{year}+\beta_{31}\texttt{surgery}\right\}, \;\;\; t > 0, 		  \label{eq:scr1} \\
	h_{13}(t \mid h_{02}, {\bm\beta}_{2}) &=& h_{02}(t)\exp\left\{\beta_{12}\texttt{age}+\beta_{22}\texttt{year}+\beta_{32}\texttt{surgery}\right\}, \;\;\; t > 0, 		  \label{eq:scr2} \\
	h_{23}(t \mid h_{03}, {\bm\beta}_{3}, T_{12}=t_{12}) &=& h_{03}(t-t_{12})\exp\left\{\beta_{13}\texttt{age}+\beta_{23}\texttt{year}+\beta_{33}\texttt{surgery}\right\}, \;\;\; t > t_{12}, \label{eq:scr3}
\end{eqnarray}

\noindent with $h_{0k}(t)=\lambda_{k} \alpha_{k} t^{\alpha_{k}-1}$ specified as a Weibull baseline hazard function, where $\alpha_{k}$ and $\lambda_{k}$ are the shape and scale parameters, respectively, for $k=1,2,3$; and ${\bm\beta}_{k}=(\beta_{1k},\beta_{2k},\beta_{3k})^{\top}$ are regression coefficients for the \texttt{age}, \texttt{year} and \texttt{surgery} covariates, respectively, We assume prior independence and specify prior marginal based on non-informative distributions commonly employed in the literature. The $\beta$'s follow a N($0,0.001$), while $\lambda$'s and $\alpha$'s follow a Gamma($0.01,0.01$) and a Un($0,10$), respectively. Note that in (\ref{eq:scr3}) we adopt the semi-Markovian specification, but it could also be the Markovian one \citep{alvares2019}.

\subsection{Model implementation}

We have created a design matrix \texttt{X} with the covariates \texttt{age}, \texttt{year} and \texttt{surgery}, and defined a \texttt{time3} $=t-t_{12}$ variable according to semi-Markovian specification:
\begin{verbatim}
R> X <- model.matrix(~ age + year + surgery, data = heart2)
R> X <- X[,-1] # Remove intercept
R> time3 <- heart2$times2
R> time3[time3 == 0] <- 0.0001
\end{verbatim}

The values of \texttt{time3} equal to zero have been replaced by 0.0001 to avoid computational problems when calculating $(t-t_{12})^{\alpha_{3}-1}$.

Listing~\ref{lst:SCR} shows a generic implementation of an illness-death model in \texttt{BUGS} syntax using \textit{heart2} data.

\begin{figure}[htb!] \small
\begin{lstlisting}[caption={Illness-death model in \texttt{BUGS} syntax (file named as \textbf{IllDeath.txt}).}, label={lst:SCR}]
model{
  for(i in 1:n){
    # Linear predictor
    elinpred[i,1] <- exp(inprod(beta[,1],X[i,]))
    elinpred[i,2] <- exp(inprod(beta[,2],X[i,]))
    elinpred[i,3] <- exp(inprod(beta[,3],X[i,]))
    # Log-hazard functions
    logHaz[i,1] <- log(lambda[1] * alpha[1] * pow(t1[i],alpha[1]-1) * elinpred[i,1])
    logHaz[i,2] <- log(lambda[2] * alpha[2] * pow(t2[i],alpha[2]-1) * elinpred[i,2])
    logHaz[i,3] <- log(lambda[3] * alpha[3] * pow(t3[i],alpha[3]-1) * elinpred[i,3])
    # Log-survival functions
    logSurv[i,1] <- -lambda[1] * pow(t1[i],alpha[1]) * elinpred[i,1]
    logSurv[i,2] <- -lambda[2] * pow(t2[i],alpha[2]) * elinpred[i,2]
    logSurv[i,3] <- -lambda[3] * pow(t3[i],alpha[3]) * elinpred[i,3]

    # Definition of the log-likelihood using zeros trick
    phi[i] <- 100000 - inprod(event[i,],logHaz[i,]) - sum(logSurv[i,])
    zeros[i] ~ dpois(phi[i])
  }

  # Prior distributions
  for(k in 1:3){
    for(l in 1:Nbetas){ beta[l,k] ~ dnorm(0,0.001) }
    lambda[k] ~ dgamma(0.01,0.01)
    alpha[k] ~ dunif(0,10)
  }
}
\end{lstlisting}
\end{figure}

Once the variables have been defined, a list with all the elements required in the model is created:
\begin{verbatim}
R> d.jags <- list(n = nrow(heart2), t1 = heart2$times1, t2 = heart2$time, t3 = time3,
+    X = X, event = event, zeros = rep(0, nrow(heart2)), Nbetas = ncol(X))
\end{verbatim}

The initial values for each illness-death model parameter are passed to \texttt{JAGS} using a function that returns a list of random values:
\begin{verbatim}
R> i.jags <- function(){
+    list(beta = matrix(rnorm(3 * ncol(X)), ncol = 3), lambda = runif(3), alpha = runif(3))
+  }
\end{verbatim}

The vector of monitored/saved parameters is:
\begin{verbatim}
R> p.jags <- c("beta", "alpha", "lambda")
\end{verbatim}

Next, the \texttt{JAGS} model is compiled:
\begin{verbatim}
R> library("rjags")
R> m5 <- jags.model(data = d.jags, file = "IllDeath.txt", inits = i.jags, n.chains = 3)
\end{verbatim}

We now run the model for 1000 burn-in simulations:
\begin{verbatim}
R> update(m5, 1000)
\end{verbatim}

Finally, the model is run for 10000 additional simulations to keep one in 10 so that a proper thinning is done:
\begin{verbatim}
R> res <- coda.samples(m5, variable.names = p.jags, n.iter = 10000, n.thin = 10)
\end{verbatim}

Similarly to the first example (Section~\ref{sec:AFT}), numerical and graphical summaries of the model parameters can be obtained using the \texttt{summary} and \texttt{densplot} functions, respectively. Gelman and Rubin's convergence diagnostic can be calculated with the \texttt{gelman.diag} function, and the \texttt{traceplot} function provides a visual way to inspect sampling behaviour and assesses mixing across chains and convergence.

Next, simulations from the three Markov chains are merged together for inference:
\begin{verbatim}
R> result <- as.mcmc(do.call(rbind, res))
\end{verbatim}

The posterior samples of each parameter are obtained by:
\begin{verbatim}
R> alpha1 <- result[,1]; alpha2 <- result[,2]; alpha3 <- result[,3]
R> beta11 <- result[,4]; beta21 <- result[,5]; beta31 <- result[,6]
R> beta12 <- result[,7]; beta22 <- result[,8]; beta32 <- result[,9]
R> beta13 <- result[,10]; beta23 <- result[,11]; beta33 <- result[,12]
R> lambda1 <- result[,13]; lambda2 <- result[,14]; lambda3 <- result[,15]
\end{verbatim}

Table~\ref{table:SCR} shows posterior summaries for the illness-death model parameters using \textit{heart2} data.
\begin{table}[htb!] \centering
\caption{Posterior summaries for the illness-death model parameters. \label{table:SCR}}
\begin{threeparttable}
\begin{tabular}{lcccccc}
\hline
Parameter &  Mean & SD & $2.5\%$ & $50\%$ & $97.5\%$ & $P(\cdot > 0 \mid \mbox{data})$ \\
\hline
\multicolumn{7}{c}{From waiting list to heart transplant ($h_{12}$)} \\
\hline
$\beta_{11}$   &  0.048  & 0.015  &  0.020  &  0.047  &  0.077  & 1.000 \\
$\beta_{21}$   &  0.004  & 0.069  & -0.130  &  0.004  &  0.140  & 0.521 \\
$\beta_{31}$   &  0.220  & 0.321  & -0.440  &  0.229  &  0.826  & 0.760 \\
$\lambda_{1}$  &  0.042  & 0.016  &  0.018  &  0.039  &  0.079  & 1.000 \\
$\alpha_{1}$   &  0.778  & 0.069  &  0.645  &  0.777  &  0.916  & 1.000 \\
\hline
\multicolumn{7}{c}{From waiting list to death ($h_{13}$)} \\
\hline
$\beta_{12}$   & -0.001  & 0.018  & -0.034  & -0.002  &  0.035  & 0.462 \\
$\beta_{22}$   & -0.243  & 0.115  & -0.476  & -0.241  & -0.027  & 0.013 \\
$\beta_{32}$   & -0.785  & 0.672  & -2.223  & -0.736  &  0.391  & 0.109 \\
$\lambda_{2}$  &  0.101  & 0.047  &  0.034  &  0.092  &  0.217  & 1.000 \\
$\alpha_{2}$   &  0.379  & 0.058  &  0.272  &  0.376  &  0.500  & 1.000 \\
\hline
\multicolumn{7}{c}{From heart transplant to death ($h_{23}$)} \\
\hline
$\beta_{13}$   &  0.055  & 0.022  &  0.014  &  0.055  &  0.099  & 0.997 \\
$\beta_{23}$   & -0.008  & 0.094  & -0.196  & -0.007  &  0.174  & 0.470 \\
$\beta_{33}$   & -0.986  & 0.469  & -1.973  & -0.961  & -0.129  & 0.011 \\
$\lambda_{3}$  &  0.034  & 0.021  &  0.008  &  0.029  &  0.087  & 1.000 \\
$\alpha_{3}$   &  0.602  & 0.074  &  0.463  &  0.598  &  0.756  & 1.000 \\
\hline
\end{tabular}
\end{threeparttable}
\end{table}

As discussed in Section~\ref{sec:MS}, the transition probabilities \eqref{eq:tp11}--\eqref{eq:tp33} are the most appropriate way to analyse the evolution of each state over time. The implementation of the posterior distribution for  these transition probabilities requires auxiliary functions, similar to the calculation of the cumulative incidence function  in Section~\ref{sec:CR}. To avoid unnecessary repetitions on how to calculate these quantities of interest, we will omit their implementation here. However, the code to reproduce Figure~\ref{fig:ms} is available in Appendix~\ref{app1}.

\begin{figure}[bt]
\centering
\includegraphics[scale=0.65]{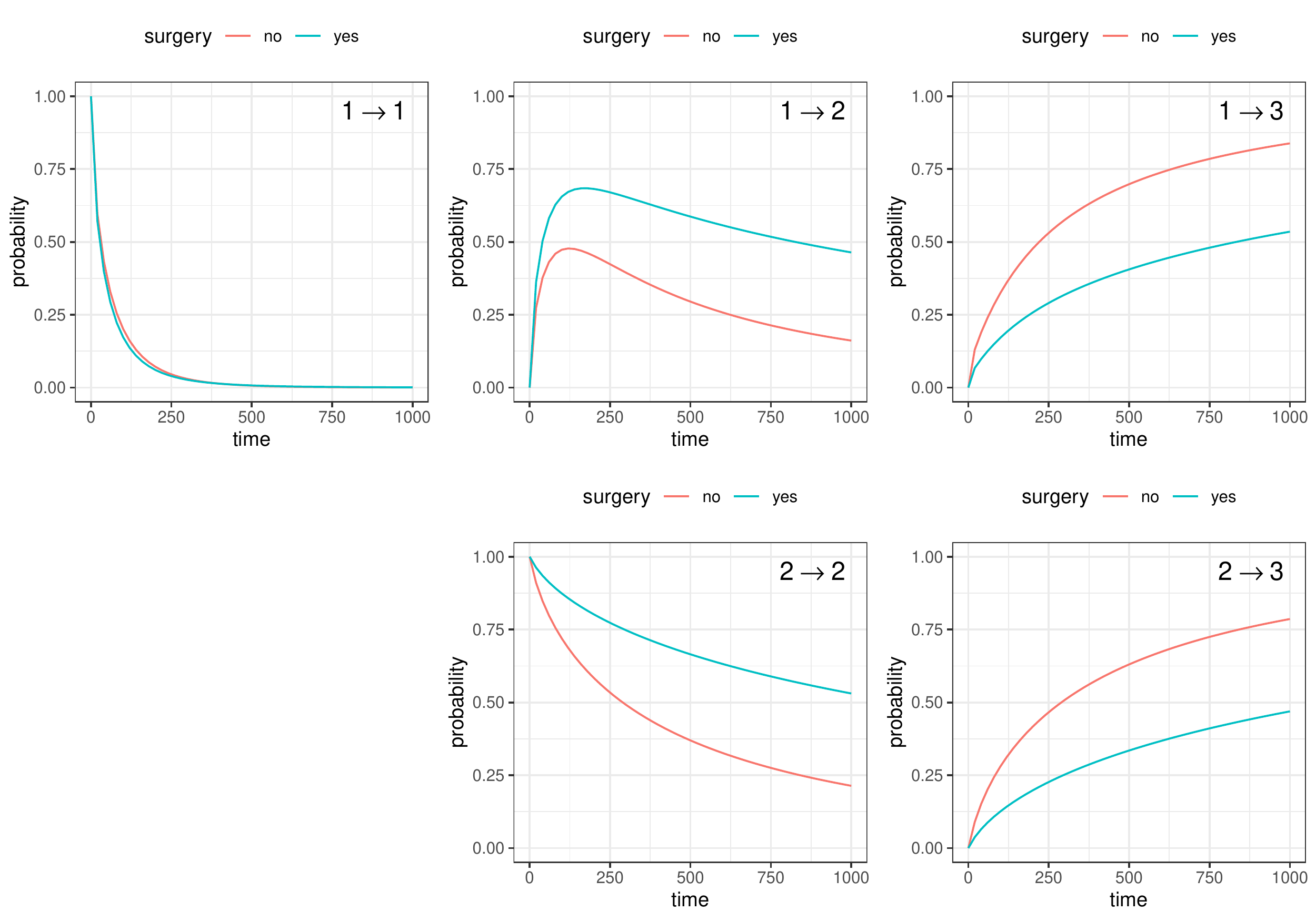}
\caption{\label{fig:ms}Posterior mean of all probability transitions from the illness-death model (\ref{eq:scr1})--(\ref{eq:scr3}) for patients with and without prior bypass surgery and median values of \texttt{age} and \texttt{year}. Graphics on the top correspond to transitions from the initial state in the waiting list. Posterior probabilities for transitions from transplant (bottom row) assume the median time $T_{12}=26$ recorded from the waiting list to heart transplant.}
\end{figure}

\section{Frailty models} \label{sec:Frailty}

Regression models include measurable covariates to improve the knowledge of the relevant failure times. However, in most survival studies, there are also individual heterogeneity that are not known or measurable. These elements are known in the statistical framework as \textit{random effects}, but in the context of survival models they are the \textit{frailty} elements \citep{aalen1994}. They can approach individual characteristics as well as heterogeneity in groups or clusters \citep{ibrahim2001}.

The most popular type of frailty models is the multiplicative shared-frailty model. It is a generalisation of the Cox regression model introduced by \cite{clayton1978} and extensively studied in \cite{hougaard2000}. Let $T_{i}$ the survival time for each individual in group $i$ with hazard function described by:
\begin{equation}
h_i(t \mid h_{0}, {\bm \beta}, w_{i})= w_{i} \, h_{0}(t) \exp\left\{{\bm x_i}^{\top}{\bm \beta}\right\}, \label{multfrailtymodel}
\end{equation}

\noindent where $w_{i}$ is the frailty term associated to group $i$. The usual probabilistic model for the frailty term is a gamma distribution with mean equal to one for identifiability purposes but also the positive stable and log-normal distributions can be considered \citep{vaupel1979}. In addition, a unity mean can be considered as a neutral frontier because frailty values greater (lower) than one increases (decreases) the individual risk. An alternative way of incorporating a frailty term in the hazard function is via an additive element as follows:
$$h_i(t \mid h_{0}, {\bm \beta}, b_{i})= h_{0}(t) \exp\left\{{\bm x_i}^{\top}{\bm \beta} + b_{i}\right\},$$

\noindent where now $b_{i}$'s are commonly assumed as normally distributed with zero mean and unknown variance.

The Bayesian framework deals with frailty models in a conceptually simpler way than the frequentist one due to the Bayesian probability conception, introduced in Section~\ref{sec1}. Hence, the inclusion of randomness through frailties in a Bayesian perspective does not add any conceptual complexity because the information regarding the risk function is expressed in probabilistic terms through its posterior distribution. Survival modelling with frailty terms is a wide issue of research that applies to all type of regression, competing risks, multivariate survival models, etc. and play a special role in joint models as we will discuss later.

\subsection{\textit{kidney} dataset}

We consider a kidney infection dataset, referred to as \textit{kidney}. It is available from the \texttt{frailtyHL} package \citep{frailtyHL}:
\begin{verbatim}
R> library("frailtyHL")
R> data("kidney")
R> str(kidney)
'data.frame':	76 obs. of  10 variables:
 $ id     : num  1 1 2 2 3 3 4 4 5 5 ...
 $ time   : num  8 16 23 13 22 28 447 318 30 12 ...
 $ status : num  1 1 1 0 1 1 1 1 1 1 ...
 $ age    : num  28 28 48 48 32 32 31 32 10 10 ...
 $ sex    : num  1 1 2 2 1 1 2 2 1 1 ...
 $ disease: Factor w/ 4 levels "Other","GN","AN",..: 1 1 2 2 1 1 1 1 1 1 ...
 $ frail  : num  2.3 2.3 1.9 1.9 1.2 1.2 0.5 0.5 1.5 1.5 ...
 $ GN     : num  0 0 1 1 0 0 0 0 0 0 ...
 $ AN     : num  0 0 0 0 0 0 0 0 0 0 ...
 $ PKD    : num  0 0 0 0 0 0 0 0 0 0 ...
\end{verbatim}

This dataset consists of times to the first and second recurrences of infection in 38 kidney patients using a portable dialysis machine. Infections can occur at the location of insertion of the catheter. The catheter is later removed if infection occurs and can be removed for other reasons, in which case the observation is censored \citep{mcgilchrist1991}. The following variables were repeated twice for each patient:
\begin{itemize}
	\item \texttt{id}: patient number.
	\item \texttt{time}: time (in days) from insertion of the catheter to infection in kidney patients using portable dialysis machine.
	\item \texttt{status}: censoring indicator (1: if patient is uncensored; 0: otherwise).
	\item \texttt{age}: age (in years) of each patient.
	\item \texttt{sex}: sex of each patient (1: if patient is male; 2: if patient is female).
	\item \texttt{disease}: disease type (GN; AN; PKD; Other).
	\item \texttt{frail}: frailty estimate from original paper.
	\item \texttt{GN}: indicator for disease type GN.
	\item \texttt{AN}: indicator for disease type AN.
	\item \texttt{PKD}: indicator for disease type PKD.
\end{itemize}

Note that the \texttt{status} variable equal to zero in any of the recurrences means that until that moment of observation there was no infection. We would also like to mention that we will use a modelling based on the \textit{clock-reset approach} \citep{kleinbaum2012}, in which time starts at zero again after each recurrence.

\subsection{Model implementation}

Time, in days, from insertion of a catheter in patient $i$th to infection is modelled trough a proportional hazard specification with multiplicative frailties:
\begin{equation}
h_{i}(t \mid h_{0}, {\bm \beta}, w_i) = w_i \, h_{0}(t) \exp\left\{\beta_{2}\texttt{sex} \right\}, \label{eq:kidney}
\end{equation}

\noindent where $w_i \sim \mbox{Gamma}(\psi,\psi)$ represents the frailty term for each individual; $h_{0}(t)=\lambda \, \alpha \, t^{\alpha-1}$ is specified as a Weibull baseline hazard function, where $\alpha$ and $\lambda=\exp(\beta_{1})$ are the shape and scale parameters, respectively; and $\beta_{2}$ is the regression coefficient for the \texttt{sex} covariate. As the purpose of this modelling is illustrative, the other covariates will not be considered. We assume prior independence and specify prior marginal based on non-informative distributions commonly employed in the literature. The $\beta$'s follow a N($0,0.001$), while $\alpha$ and $\psi$ follow a Un($0,10$) and a Gamma($0.01,0.01$), respectively.

Our variable of interest is \texttt{time}, which represents the time from insertion of the catheter to infection. The \texttt{status} covariate plays an important role in the codification of the survival and censoring times:
\begin{verbatim}
R> # Number of patients and catheters
R> n <- length(unique(kidney$id))
R> J <- 2
R> # Survival and censoring times
R> time <- kidney$time
R> cens <- time
R> time[kidney$status == 0] <- NA # Censored
R> is.censored <- as.numeric(is.na(time))
R> # Matrix format
R> time <- matrix(time, n, J, byrow = TRUE)
R> cens <- matrix(cens, n, J, byrow = TRUE)
R> is.censored <- matrix(is.censored, n, J, byrow = TRUE)
\end{verbatim}

Without loss of generality, we have created a design matrix \texttt{X} with the \texttt{sex} covariate:
\begin{verbatim}
R> sex <- kidney$sex[seq(1, 2 * n, 2)] - 1 # Reference = male
R> X <- model.matrix(~ sex)
\end{verbatim}

Listing~\ref{lst:Frailty} shows a generic implementation of a frailty model in \texttt{BUGS} syntax using \textit{kidney} data.

\begin{figure}[htb!] \small
\begin{lstlisting}[caption={Frailty model in \texttt{BUGS} syntax (file named as \textbf{Frailty.txt}).}, label={lst:Frailty}]
model{
  for(i in 1:n){
    for(j in 1:J){
      # Survival and censoring times
      is.censored[i,j] ~ dinterval(time[i,j],cens[i,j])
      time[i,j] ~ dweib(alpha,lbd[i,j])
      log(lbd[i,j]) <- inprod(beta[],X[i,]) + log(w[i])
    }
    # Multiplicative frailties
    w[i] ~ dgamma(psi,psi)
  }
	
  # Prior distributions
  for(l in 1:Nbetas){ beta[l] ~ dnorm(0.0,0.001) }
  alpha ~ dunif(0,10)
  psi ~ dgamma(0.01,0.01)

  # Derived quantity
  lambda <- exp(beta[1])
}
\end{lstlisting}
\end{figure}

\subsection{Model estimation: \texttt{JAGS} from R}

Once the variables have been defined, a list with all the elements required in the model is created:
\begin{verbatim}
R> d.jags <- list(n = n, J = J, time = time, cens = cens, X = X,
+   is.censored = is.censored, Nbetas = ncol(X))
\end{verbatim}

The initial values for each frailty model parameter are passed to \texttt{JAGS} using a function that returns a list of random values:
\begin{verbatim}
R> i.jags <- function(){ list(beta = rnorm(ncol(X)), alpha = runif(1), psi = runif(1)) }
\end{verbatim}

The vector of monitored/saved parameters is:
\begin{verbatim}
R> p.jags <- c("beta", "alpha", "lambda", "psi", "w")
\end{verbatim}

Next, the \texttt{JAGS} model is compiled:
\begin{verbatim}
R> library("rjags")
R> m6 <- jags.model(data = d.jags, file = "Frailty.txt", inits = i.jags, n.chains = 3)
\end{verbatim}

We now run the model for 10000 burn-in simulations:
\begin{verbatim}
R> update(m6, 10000)
\end{verbatim}

Finally, the model is run for 100000 additional simulations to keep one in 100 so that a proper thinning is done:
\begin{verbatim}
R> res <- coda.samples(m6, variable.names = p.jags, n.iter = 100000, thin = 100)
\end{verbatim}

Similarly to the first example (Section~\ref{sec:AFT}), numerical and graphical summaries of the model parameters can be obtained using the \texttt{summary} and \texttt{densplot} functions, respectively. Gelman and Rubin's convergence diagnostic can be calculated with the \texttt{gelman.diag} function, and the \texttt{traceplot} function provides a visual way to inspect sampling behaviour and assesses mixing across chains and convergence.

Next, simulations from the three Markov chains are merged together for inference:
\begin{verbatim}
R> result <- as.mcmc(do.call(rbind, res))
\end{verbatim}

The posterior samples of each parameter are obtained by:
\begin{verbatim}
R> alpha <- result[,1]; beta2 <- result[,3]; lambda <- result[,4]
R> psi <- result[,5]; w <- result[,6:ncol(result)]
\end{verbatim}

Table~\ref{table:Frailty} shows posterior summaries for the frailty model parameters using \textit{kidney}.
\begin{table}[htb!] \centering
\caption{Posterior summaries for the frailty model parameters. \label{table:Frailty}}
\begin{threeparttable}
\begin{tabular}{lcccccc}
\hline 
Parameter &  Mean & SD & $2.5\%$ & $50\%$ & $97.5\%$ & $P(\cdot > 0 \mid \mbox{data})$ \\
\hline
$\beta_{2}$ (\texttt{sex})  &  -1.908  &  0.555  &  -3.064  & -1.889  & -0.876  & 0.000 \\
$\alpha$                    &   1.233  &  0.167  &   0.929  &  1.222  &  1.592  & 1.000 \\
$\lambda$                   &   0.019  &  0.012  &   0.004  &  0.017  &  0.050  & 1.000 \\
$\psi$                      &   2.417  &  2.101  &   0.779  &  1.878  &  7.844  & 1.000 \\
\hline
\end{tabular}
\end{threeparttable}
\end{table}

\begin{figure}[bt]
\centering
\includegraphics[scale=0.6]{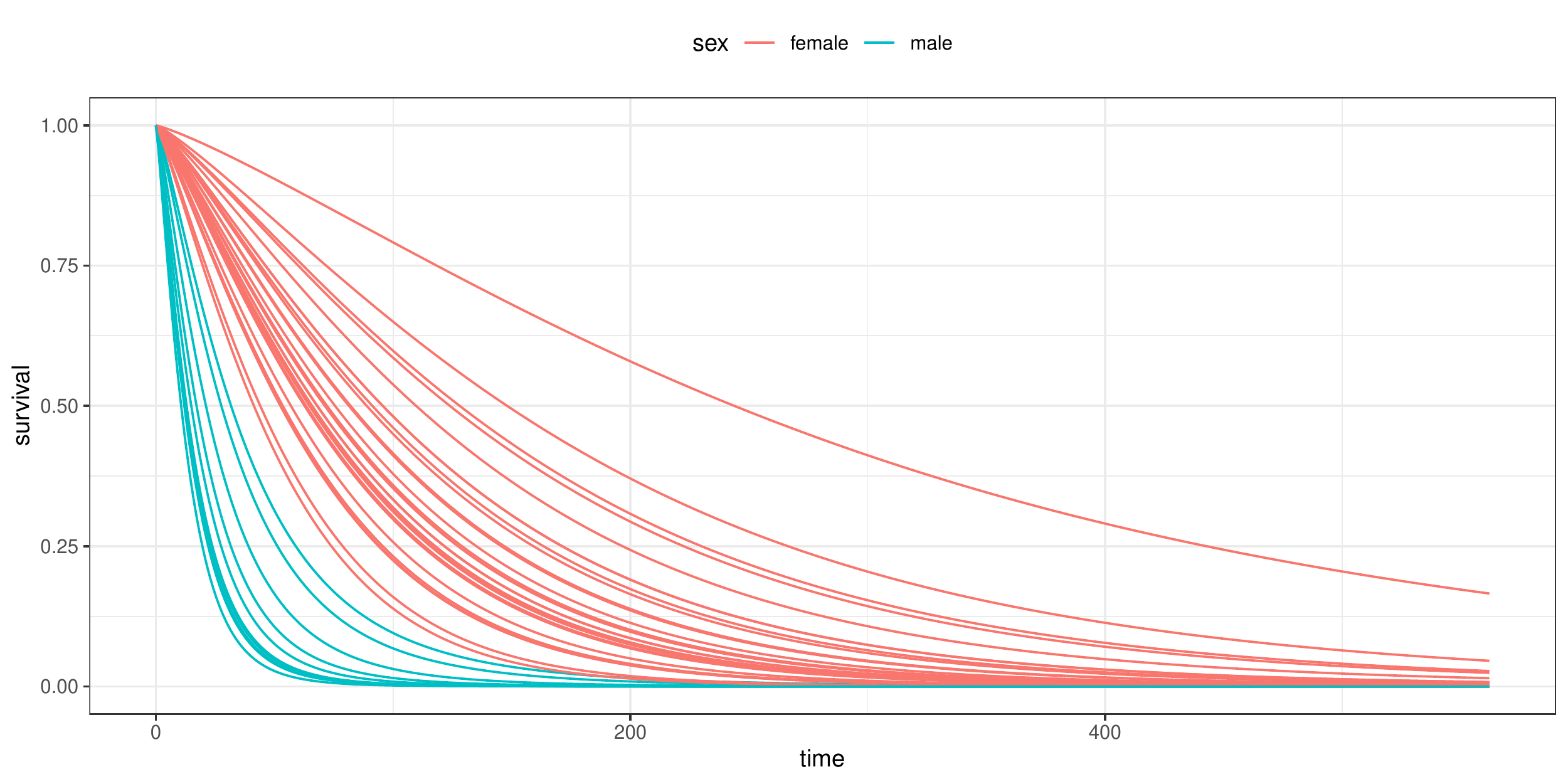}
\caption{\label{fig:frailty}Posterior mean of the individual survival function from the frailty model (\ref{eq:kidney}).}
\end{figure}

The individual survival curve based on posterior samples is a relevant information in this type of studies. So, we can summarise the posterior distribution of the individual survival curve in a grid of points as follows:
\begin{verbatim}
R> grid <- 1000
R> time <- seq(0, max(kidney$time), len = grid)
R> surv <- matrix(NA, n, grid)
R> for(i in 1:n){
+    for(k in 1:grid){
+       surv[i, k] <- mean(exp(-w[i] * lambda * exp(beta2 * sex[i]) * time[k]^alpha))
+    }
+  }
\end{verbatim}

Next, we can differentiate the survival curves by sex (code below). Figure~\ref{fig:frailty} shows such curves for all patients in the \textit{kidney} data.
\begin{verbatim}
R> library("ggplot2")
R> sex.col <- sex
R> sex.col[sex == 0] <- "male"
R> sex.col[sex == 1] <- "female"
R> df <- data.frame(time = rep(time, n), survival = c(t(surv)),
+   patient = rep(1:n, each = grid), sex = rep(sex.col, each = grid))
R> ggplot(data = df, aes(x = time, y = survival, group = patient, colour = sex)) +
+   geom_line() + theme_bw() + theme(legend.position = "top")
\end{verbatim}

The curves shown in Figure~\ref{fig:frailty} represent the posterior mean of the probability of infection from any catheter insertion at each time considering the two replicates per patient.

\section{Joint models of longitudinal and survival data} \label{sec:JM}

Joint modelling of longitudinal and time-to-event data is an increasingly productive area of statistical research that examines the association between longitudinal and survival processes \citep{rizopoulos2012}. It enhances survival modelling with the inclusion of internal time-dependent covariates as well as longitudinal modelling by allowing for the inclusion of non-ignorable dropout mechanisms through survival tools. Joint models were introduced during the 90s \citep{gruttola1994, tsiatis1995, faucett1996, wulfsohn1997} and since then, have been applied to a great variety of studies mainly in epidemiological and biomedical areas.

Bayesian joint models assume a full joint distribution for the longitudinal ($y$) and the survival processes ($s$) as well as the subject-specific random effects vector ($\bm b$) and the parameters and hyperparameters ($\bm \theta$) of the model \citep{armero2018}. Usually, they can be defined as follow:
\begin{equation}
f(y, s, {\bm b}, {\bm \theta}) = f(y, s \mid {\bm b}, {\bm \theta}) \, f({\bm b} \mid {\bm \theta}) \, \pi({\bm \theta}), \label{jointmodel}
\end{equation}

\noindent which factorises as the product of the joint conditional distribution $f(y, s \mid {\bm b}, {\bm \theta})$, the conditional distribution $f({\bm b} \mid {\bm \theta})$ of the random effects, and the prior distribution $\pi({\bm \theta})$. There are different proposals for the specification of the conditional distribution $f(y, s \mid {\bm b}, {\bm \theta})$. The most popular approaches are the share-parameter models and the joint latent class models.

Shared-parameter models are a type of joint models where the longitudinal and time-to-event processes are connected by means of a common set of subject-specific random effects. These models make possible to quantify both the population and individual effects of the underlying longitudinal outcome on the risk of an event and allow to obtain individualised time-dynamic predictions \citep{wu1988, hogan1997, hogan1998}. In particular, this approach postulates conditional independence between the longitudinal and survival processes given the random effects and the parameters: $$f(y, s \mid {\bm b}, {\bm \theta}) = f(y \mid {\bm b}, {\bm \theta}) f(s \mid {\bm b}, {\bm \theta}).$$

The joint latent class model is based on finite mixtures \citep{proust-Lima2014}. Heterogeneity among the individuals is classified into a finite number $G$ of homogeneous latent clusters which share the same longitudinal trajectory and the same risk function. Both elements are also conditionally independent within the subsequent latent group as follows: $$f(y, s \mid L=g, {\bm b}, {\bm \theta}) = f(y \mid L=g, {\bm b}, {\bm \theta}) f(s \mid L=g, {\bm \theta}),$$

\noindent where $L$ is a random variable that measures the uncertainty on the membership of each individual to each group, usually modelled by means of a multinomial logistic model. These models are possibly the most complex. Predicting observations from models with random effects is not easy. In this case, predicting longitudinal observations will be complicated because the conditional marginal distribution $f(y \mid L=g, {\bm b}, {\bm \theta})$ necessary to obtain the corresponding predictive posterior  distribution depends on both random effects and latent groups. On the other hand, the prediction of survival times does seem to be computationally simpler because the conditional distribution $f(s \mid L=g, {\bm \theta})$ does not depend on individual random effects.
 
All these proposals account for a particular type of conditional independence between the longitudinal and the survival processes which facilitates the modelling into longitudinal and survival submodels with various types of connectors. This general structure allows any type of modelling for the survival process such as frailty survival regression models, competing risks with frailties, cure models with frailties as well as linear mixed models or generalised linear mixed models for the longitudinal process \citep{taylor2013, rizopoulos2015, armero2016, rue2017}. See \cite{armero2020} for a short review on Bayesian joint models up to date.

\subsection{\textit{prothro} dataset}

We consider a liver cirrhosis dataset, referred to as \textit{prothro} (longitudinal information) and \textit{prothros} (survival information). It is available from the \texttt{JMbayes} package \citep{JMbayes}:
\begin{verbatim}
R> library("JMbayes")
R> data("prothro")
R> data("prothros")
R> str(prothro); str(prothros)
'data.frame':	2968 obs. of  9 variables:
 $ id   : num  1 1 1 2 2 2 2 2 2 2 ...
 $ pro  : num  38 31 27 51 73 90 64 54 58 90 ...
 $ time : num  0 0.244 0.381 0 0.687 ...
 $ treat: Factor w/ 2 levels "placebo","prednisone": 2 2 2 2 2 2 2 2 2 2 ...
 $ Time : num  0.413 0.413 0.413 6.754 6.754 ...
 $ start: num  0 0.244 0.381 0 0.687 ...
 $ stop : num  0.244 0.381 0.413 0.687 0.961 ...
 $ death: num  1 1 1 1 1 1 1 1 1 1 ...
 $ event: num  0 0 1 0 0 0 0 0 0 0 ...
'data.frame':	488 obs. of  4 variables:
 $ id   : num  1 2 3 4 5 6 7 8 9 10 ...
 $ Time : num  0.413 6.754 13.394 0.794 0.75 ...
 $ death: num  1 1 0 1 1 1 1 0 0 1 ...
 $ treat: Factor w/ 2 levels "placebo","prednisone": 2 2 2 2 2 2 1 1 1 1 ...
\end{verbatim}

These datasets are part of a placebo-controlled randomised trial on 488 liver cirrhosis patients, where the longitudinal observations of a biomarker (prothrombin) are recorded \citep{andersen1993}. For our illustrative purpose, only the following variables are relevant:
\begin{itemize}
	\item \texttt{id}: patient number.
	\item \texttt{pro}: prothrombin measurements.
	\item \texttt{time}: time points at which the prothrombin measurements were taken.
	\item \texttt{treat}: randomised treatment (placebo or prednisone).
	\item \texttt{Time}: time (in years) from the start of treatment until death or censoring.	
	\item \texttt{death}: censoring indicator (1: if patient is died; 0: otherwise).	
\end{itemize}

\subsection{Model implementation}

We assume a shared-parameter model specified by a linear mixed-effects model \citep{laird1982} for the longitudinal response and a Cox model for the survival part. The linear mixed-effects model which describes the subject-specific prothrombin evolution of individual $i$ over time is given by:
\begin{equation}
y_i(t) = \beta_{L1} + b_{i1} + (\beta_{L2} + b_{i2}) \, t + \beta_{L3}\texttt{treat} + \epsilon_i(t), \;\;\; i=1,\ldots,n, \label{eq:JM1}
\end{equation}

\noindent where $y_i(t)$ represents the prothrombin value at time $t$ for individual $i$; $\beta_{L1}$ and $\beta_{L2}$ are fixed effects for intercept and slope, respectively, with $b_{i1}$ and $b_{i2}$ being the respective individual random effects; $\beta_{L3}$ is the regression coefficient for the \texttt{treat} covariate; and $\epsilon_i(t)$ is a measurement error for individual $i$ at time $t$. We assume that the individual random effects, ${\bm b_i} = (b_{i1},b_{i2})^{\top}$, given $\Sigma$ follow a joint bivariate normal distribution with mean vector $(0,0)^{\top}$ and variance-covariance matrix $\Sigma$, and that the errors are conditionally i.i.d. as $(\epsilon_i(t) \mid \sigma^{2}) \sim \mbox{N}(0,\sigma^{2})$, where $\sigma^{2}$ represents the error variance. Random effects and error terms were assumed mutually independent.

Survival time for individual $i$ is modelled from a proportional hazard specification which includes in the exponential term the random effects $b_{i1}$ and $b_{i2}$ in  (\ref{eq:JM1}) as follows:
\begin{equation}
h_i(t \mid h_{0}, {\bm \beta}_{S}, \gamma, {\bm b_i}) = h_{0}(t) \exp\left\{\beta_{S2}\texttt{treat} + \gamma(b_{i1} + b_{i2}t) \right\}, \label{eq:JM2}
\end{equation}

\noindent with $h_{0}(t)=\lambda \, \alpha \, t^{\alpha-1}$ specified as a Weibull baseline hazard function, where $\alpha$ and $\lambda=\exp(\beta_{S1})$ are the shape and scale parameters, respectively; $\beta_{S2}$ is the regression coefficient for the \texttt{treat} covariate; and $\gamma$ is an association parameter that measure the strength of the link between the random effects associated to individual $i$ of the longitudinal submodel and their risk of death at time $t$.

We assume prior independence and specify prior marginal distributions based on non-informative distributions commonly employed in the literature. $\beta_{L}$'s, $\beta_{S}$' and $\gamma$ follow a N($0,0.001$), while $\alpha$ and $\sigma$ follow a Un($0,10$) and a Un($0,100$), respectively, and $\Sigma$ follows an Inv-Wishart($V,r$), where $V$ is a 2$\times$2 identity matrix and $r=2$ is the degrees-of-freedom parameter. The position of the inverse-Wishart distribution in parameter space is specified by $V$, while $r$ set the certainty about the prior information in the scale matrix \citep{schuurman2016}. The larger the $r$, the higher the certainty about the information in $V$, and the more informative is the distribution. Hence, in our application, the least informative specification then results when $r=2$ (number of random effects), which is the lowest possible number of $r$. Additionally, $V$ as an identity matrix has the appealing feature that each of the correlations in $\Sigma$ has, marginally, a uniform prior distribution \citep{bda2013}.

Our variable of interest is \texttt{Time} (\textit{prothros} file), which represents the time from the start of treatment until death or censoring. The \texttt{death}=1 variable indicates an uncensored time:
\begin{verbatim}
R> # Number of patients and number of longitudinal observations per patient
R> n <- nrow(prothros)
R> M <- table(prothro$id)
R> # Survival and censoring times
R> Time <- prothros$Time
R> death <- prothros$death
\end{verbatim}

The (log)prothrombin observations and their respective measurement times (\textit{prothro} file) have been rearranged in matrix format:
\begin{verbatim}
R> # Longitudinal information in matrix format
R> time <- matrix(NA, n, max(M))
R> log.proth <- matrix(NA, n, max(M))
R> count <- 1
R> for(i in 1:n){
+     log.proth[i, 1:M[i]] <- log(prothro$pro[count:(M[i] + count - 1)])
+     time[i, 1:M[i]] <- prothro$time[count:(M[i] + count - 1)]
+     count <- count + M[i]
+  }
\end{verbatim}

We have created the survival design matrix composed of an intercept and a treatment variable:
\begin{verbatim}
R> treat <- as.numeric(prothros$treat) - 1 # Reference = placebo
R> XS <- model.matrix(~ treat) # Fixed effects
\end{verbatim}

We have split the longitudinal design matrix into two parts, \texttt{XL} (fixed effects) and \texttt{ZL} (random effects):
\begin{verbatim}
R> XL <- array(1, dim = c(n, max(M), 3)) # Fixed effects
R> XL[, , 2] <- time; XL[, , 3] <- treat
R> ZL <- array(1, dim = c(n, max(M), 2)) # Random effects
R> ZL[, , 2] <- time
\end{verbatim}

The survival function for individual $i$, $S_i(t \mid h_{0}, {\bm \beta}_{S}, \gamma, {\bm b_i})=\exp\left\{-\int_{0}^{t}\,h_i(u \mid h_{0}, {\bm \beta}_{S}, \gamma, {\bm b_i}) \,\mbox{d}u\right\}$can be efficiently approximated using some Gaussian quadrature method, available from the \texttt{statmod} package \citep{statmod}. For our analysis, we have used 15-point Gauss-Legendre quadrature rule, as is done in \cite{armero2018}:
\begin{verbatim}
R> # Gauss-Legendre quadrature (15 points)
R> library("statmod")
R> glq <- gauss.quad(15, kind = "legendre")
R> xk <- glq$nodes   # Nodes
R> wk <- glq$weights # Weights
R> K <- length(xk)   # K-points
\end{verbatim}

Listing~\ref{lst:JM} shows a generic implementation of a joint model in \texttt{BUGS} syntax using \textit{prothro}/\textit{prothros} data.

\begin{figure}[htb!] \small
\begin{lstlisting}[caption={Joint model in \texttt{BUGS} syntax (file named as \textbf{JM.txt}).}, label={lst:JM}]
model{
  for(i in 1:n){
    # Longitudinal observations
    for(j in 1:M[i]){
       log.proth[i,j] ~ dnorm(mu[i,j],tau)
   	   mu[i,j] <- inprod(betaL[],XL[i,j,]) + inprod(b[i,],ZL[i,j,])
    }

    # Survival and censoring times
    # Hazard function at integration points
    for(j in 1:K){
       haz[i,j] <- alpha * pow(Time[i]/2*(xk[j]+1),alpha-1) *
                   exp(inprod(betaS[],XS[i,])+gamma*(b[i,1]+b[i,2]*(Time[i]/2*(xk[j]+1))))
    }
    # Log-survival function with Gauss-Legendre quadrature
    logSurv[i] <- -Time[i]/2 * inprod(wk, haz[i,])

    # Definition of the survival log-likelihood using zeros trick
    phi[i] <- 100000 - death[i] * log(haz[i,K]) - logSurv[i]
    zeros[i] ~ dpois(phi[i])

    # Random effects
    b[i,1:Nb] ~ dmnorm(mub[],Omega[,])
  }

  # Prior distributions
  for(l in 1:NbetasS){ betaS[l] ~ dnorm(0.0,0.001) }
  gamma ~ dnorm(0.0,0.001)
  alpha ~ dunif(0,10)
  for(l in 1:NbetasL){ betaL[l] ~ dnorm(0.0,0.001) }
  tau <- pow(sigma,-2)
  sigma ~ dunif(0,100)
  Omega[1:Nb,1:Nb] ~ dwish(V[,],Nb)
  Sigma[1:Nb,1:Nb] <- inverse(Omega[,])

  # Derived quantity
  lambda <- exp(betaS[1])
}
\end{lstlisting}
\end{figure}

Once the variables have been defined, a list with all the elements required in the model is created:
\begin{verbatim}
R> d.jags <- list(n = n, M = M, Time = Time, XS = XS, log.proth = log.proth, XL = XL, ZL = ZL,
+    XS = XS, death = death, mub = rep(0, 2), V = diag(1, 2), Nb = 2, zeros = rep(0, n),
+    NbetasL = dim(XL)[3], NbetasS = ncol(XS), K = length(xk), xk = xk, wk = wk)
\end{verbatim}

The variables \texttt{mub} and \texttt{V} represent, respectively, the mean of the random effects normally distributed and the scale matrix of the Wishart distribution which models the precision of the random effects.

The initial values for each joint model parameter are passed to \texttt{JAGS} using a function that returns a list of random values:
\begin{verbatim}
R> i.jags <- function(){
+    list(betaS = rnorm(ncol(XS)), gamma = rnorm(1), alpha = runif(1),
+      betaL = rnorm(dim(XL)[3]), sigma = runif(1), Omega = diag(runif(2)))
+  }
\end{verbatim}

The vector of monitored/saved parameters is:
\begin{verbatim}
R> p.jags <- c("betaS", "gamma", "alpha", "lambda", "betaL", "sigma", "Sigma", "b")
\end{verbatim}

Next, the \texttt{JAGS} model is compiled:
\begin{verbatim}
R> library("rjags")
R> m7 <- jags.model(data = d.jags, file = "JM.txt", inits = i.jags, n.chains = 3)
\end{verbatim}

We now run the model for 1000 burn-in simulations:
\begin{verbatim}
R> update(m7, 1000)
\end{verbatim}

Finally, the model is run for 10000 additional simulations to keep one in 10 so that a proper thinning is done:
\begin{verbatim}
R> res <- coda.samples(m7, variable.names = p.jags, n.iter = 10000, thin = 10)
\end{verbatim}

Similarly to the first example in Section~\ref{sec:AFT}, numerical and graphical summaries of the model parameters can be obtained using the \texttt{summary} and \texttt{densplot} functions, respectively. Gelman and Rubin's convergence diagnostic can be calculated with the \texttt{gelman.diag} function, and the \texttt{traceplot} function provides a visual way to inspect sampling behaviour and assesses mixing across chains and convergence.

Next, simulations from the three Markov chains are merged together for inference:
\begin{verbatim}
R> result <- as.mcmc(do.call(rbind, res))
\end{verbatim}

The posterior samples of each parameter are obtained by:
\begin{verbatim}
R> Sigma2.11 <- result[,1]; Sigma2.12 <- result[,2]; Sigma2.22 <- result[,4]
R> alpha <- result[,5]; b1 <- result[,6:(n+5)]; b2 <- result[,(n+6):(2*n+5)]
R> betaL1 <- result[,(2*n+6)]; betaL2 <- result[,(2*n+7)]; betaL3 <- result[,(2*n+8)]
R> betaS2 <- result[,(2*n+10)]; gamma <- result[,(2*n+11)]
R> lambda <- result[,(2*n+12)]; sigma <- result[,(2*n+13)]
\end{verbatim}

Table~\ref{table:JM} shows posterior summaries for the joint model parameters using \textit{prothro}/\textit{prothros} data.
\begin{table}[htb!] \centering
\caption{Posterior summaries for the joint model parameters. \label{table:JM}}
\begin{threeparttable}
\begin{tabular}{lcccccc}
\hline 
Parameter &  Mean & SD & $2.5\%$ & $50\%$ & $97.5\%$ & $P(\cdot > 0 \mid \mbox{data})$ \\
\hline
$\beta_{S2}$ (\texttt{treat})     &  0.073  & 0.138  & -0.191  &  0.071  &  0.343  & 0.703 \\
$\gamma$ (\texttt{assoc})         & -2.269  & 0.180  & -2.640  & -2.264  & -1.923  & 0.000 \\
$\lambda$                         &  0.187  & 0.023  &  0.145  &  0.186  &  0.233  & 1.000 \\
$\alpha$                          &  0.934  & 0.049  &  0.841  &  0.934  &  1.034  & 1.000 \\
\hline
$\beta_{L1}$ (\texttt{intercept}) &  4.276  & 0.021  &  4.235  &  4.276  &  4.318  & 1.000 \\
$\beta_{L2}$ (\texttt{slope})     & -0.004  & 0.007  & -0.018  & -0.004  & -0.010  & 0.301 \\
$\beta_{L3}$ (\texttt{treat})     & -0.099  & 0.030  & -0.159  & -0.098  & -0.040  & 0.000 \\
$\sigma$                          &  0.258  & 0.004  &  0.250  &  0.258  &  0.265  & 1.000 \\
\hline
$\Sigma_{11}$                     &  0.098  & 0.008  &  0.083  &  0.098  &  0.116  & 1.000 \\
$\Sigma_{22}$                     &  0.013  & 0.001  &  0.011  &  0.013  &  0.017  & 1.000 \\
$\Sigma_{12}$                     & -0.003  & 0.003  & -0.009  & -0.003  &  0.003  & 0.145 \\
\hline
\end{tabular}
\end{threeparttable}
\end{table}

Most of the parameters are interpreted similarly to the ones in  previous examples. However, the association parameter, $\gamma$, plays an important role in this type of models. In our illustration, the posterior mean of $\gamma$ is negative, -2.269, and $P(\gamma > 0 \mid \mbox{data}) = 0$, indicating a strong negative association of the prothrombin measurements with respect to vital status. In other words, a negative value for $\gamma$ means that low values or decreasing trends of prothrombin increase the risk of death.

\section{Conclusions} \label{sec:conclusions}

The analysis of time until an event of interest requires a suitable and flexible modelling and has applications in several applied fields. The \texttt{BUGS} language offers the opportunity of easily use and adapt Bayesian hierarchical models without the need to manually implement Markov chain Monte Carlo methods. Hence, this paper has summarised some of the most popular survival models and has focused on the Bayesian paradigm to make the
inferential procedure. Furthermore, for each of the models proposed we have provided the codes in \texttt{BUGS} syntax, so that model can be fit with the support of the \texttt{rjags} package from the R language.

We have discussed all the implementation details of the following Bayesian survival models: accelerated failure time, proportional hazards, mixture cure, competing risks, multi-state, frailty, and joint models of longitudinal and survival data. Moreover, the computation of quantities of interest derived from posterior samples has been provided as well as some graphs that assist in the interpretation of results and decision making. The paper has also briefly presented other Bayesian R-packages that handle time-to-event data. 

In conclusion, we hope this paper will encourage researchers who use survival models make their analyses based on the Bayesian paradigm from the \texttt{BUGS} codes we have provided and easily adapt them to novel settings. In addition, the descriptions of the survival models provided herein could also be used as a guidance to implement these models using other similar languages such as, for example, \texttt{Stan} \citep{Stan}.

\section*{Acknowledgments}

This work was partially supported by FONDECYT (Chile), Grant Number: 11190018; Consejer{\'i}a de Educaci{\'o}n, Cultura y Deportes (JCCM, Spain), Grant Number: PPIC-2014-001-P; FEDER, Grant Number: SBPLY/17/180501/000491; and Ministerio de Ciencia e Innovaci{\'o}n (MCI, Spain), Grant Number: PID2019-106341GB-I00.

\bibliographystyle{plainnat}
\bibliography{refs}

\begin{thebibliography}{85}
\providecommand{\natexlab}[1]{#1}
\providecommand{\url}[1]{\texttt{#1}}
\expandafter\ifx\csname urlstyle\endcsname\relax
  \providecommand{\doi}[1]{doi: #1}\else
  \providecommand{\doi}{doi: \begingroup \urlstyle{rm}\Url}\fi

\bibitem[Aalen(1994)]{aalen1994}
O.~O. Aalen.
\newblock {Effects of Frailty in Survival Analysis}.
\newblock \emph{Statistical Methods in Medical Research}, 3\penalty0
  (3):\penalty0 227--243, 1994.
\newblock \doi{10.1177/096228029400300303}.

\bibitem[Alvares et~al.(2019)Alvares, Haneuse, Lee, and Lee]{alvares2019}
D.~Alvares, S.~Haneuse, C.~Lee, and K.~H. Lee.
\newblock {SemiCompRisks: An R Package for the Analysis of Independent and
  Cluster-correlated Semi-competing Risks Data}.
\newblock \emph{The R Journal}, 11\penalty0 (1):\penalty0 376--400, 2019.
\newblock \doi{10.32614/rj-2019-038}.

\bibitem[Andersen and Keiding(2002)]{andersen2002}
P.~K. Andersen and N.~Keiding.
\newblock {Multi-state Models for Event History Analysis}.
\newblock \emph{Statistical Methods in Medical Research}, 11\penalty0
  (2):\penalty0 91--115, 2002.
\newblock \doi{10.1191/0962280202sm276ra}.

\bibitem[Andersen and Perme(2008)]{andersen2008}
P.~K. Andersen and M.~P. Perme.
\newblock {Inference for Outcome Probabilities in Multi-state Models}.
\newblock \emph{Lifetime Data Analysis}, 14\penalty0 (4):\penalty0 405--431,
  2008.
\newblock \doi{10.1007/s10985-008-9097-x}.

\bibitem[Andersen et~al.(1993)Andersen, Borgan, Gill, and
  Keiding]{andersen1993}
P.~K. Andersen, {\O}.~Borgan, R.~D. Gill, and N.~Keiding.
\newblock \emph{{Statistical Models Based on Counting Processes}}.
\newblock Springer, 1st edition, 1993.

\bibitem[Armero(2020)]{armero2020}
C.~Armero.
\newblock {Bayesian Joint Models for Longitudinal and Survival Data}.
\newblock \emph{Wiley StatsRef: Statistics Reference Online}, 2020.
\newblock \doi{arXiv:2005.12822}.

\bibitem[Armero et~al.(2016{\natexlab{a}})Armero, Cabras, Castellanos, Perra,
  Quir{\'o}s, Oruez{\'a}bal, and S{\'a}nchez-Rubio]{armero2016b}
C.~Armero, S.~Cabras, M.~E. Castellanos, S.~Perra, A.~Quir{\'o}s, M.~J.
  Oruez{\'a}bal, and J.~S{\'a}nchez-Rubio.
\newblock {Bayesian Analysis of a Disability Model for Lung Cancer Survival}.
\newblock \emph{Statistics Methods in Medical Research}, 25\penalty0
  (1):\penalty0 336--351, 2016{\natexlab{a}}.
\newblock \doi{10.1177/0962280212452803}.

\bibitem[Armero et~al.(2016{\natexlab{b}})Armero, Forn{\'e}, Ru{\'{e}}, Forte,
  Perpi{\~n}{\'a}n, G{\'o}mez, and Bar{\'e}]{armero2016}
C.~Armero, C.~Forn{\'e}, M.~Ru{\'{e}}, A.~Forte, H.~Perpi{\~n}{\'a}n,
  G.~G{\'o}mez, and M.~Bar{\'e}.
\newblock {Bayesian Joint Ordinal and Survival Modeling for Breast Cancer Risk
  Assessment}.
\newblock \emph{Statistics in Medicine}, 35\penalty0 (28):\penalty0 5267--5282,
  2016{\natexlab{b}}.
\newblock \doi{10.1002/sim.7065}.

\bibitem[Armero et~al.(2018)Armero, Forte, Perpi{\~n}{\'a}n, Sanahuja, and
  Agust{\'i}]{armero2018}
C.~Armero, A.~Forte, H.~Perpi{\~n}{\'a}n, M.~J. Sanahuja, and S.~Agust{\'i}.
\newblock {Bayesian Joint Modeling for Assessing the Progression of Chronic
  Kidney Disease in Children}.
\newblock \emph{Statistical Methods in Medical Research}, 27\penalty0
  (1):\penalty0 298--311, 2018.
\newblock \doi{10.1177/0962280216628560}.

\bibitem[Auguie and Antonov(2017)]{gridextra}
B.~Auguie and A.~Antonov.
\newblock \emph{{\texttt{gridExtra}: Miscellaneous Functions for ``Grid''
  Graphics}}.
\newblock R package version 2.3,
  \url{https://CRAN.R-project.org/package=gridExtra}, 2017.

\bibitem[Berkson and Gage(1952)]{berkson1952}
J.~Berkson and R.~P. Gage.
\newblock {Survival Curve for Cancer Patients Following Treatment}.
\newblock \emph{Journal of the American Statistical Association}, 47\penalty0
  (259):\penalty0 501--515, 1952.
\newblock \doi{10.1080/01621459.1952.10501187}.

\bibitem[Beyersmann et~al.(2007)Beyersmann, Dettenkofer, Bertz, and
  Schumacher]{beyersmann2007}
J.~Beyersmann, M.~Dettenkofer, H.~Bertz, and M.~Schumacher.
\newblock {A Competing Risks Analysis of Bloodstream Infection After Stem-Cell
  Transplantation Using Subdistribution Hazards and Cause-Specific Hazards}.
\newblock \emph{Statistics in Medicine}, 26\penalty0 (30):\penalty0 5360--5369,
  2007.
\newblock \doi{10.1002/sim.3006}.

\bibitem[Bogaerts et~al.(2017)Bogaerts, Kom{\'a}rek, and
  Lesaffre]{bogaerts2017}
K.~Bogaerts, A.~Kom{\'a}rek, and E.~Lesaffre.
\newblock \emph{{Survival Analysis with Interval-Censored Data: A Practical
  Approach with Examples in R, SAS, and BUGS}}.
\newblock Chapman \& Hall/CRC, 1st edition, 2017.

\bibitem[Borchers(2019)]{pracma}
H.~W. Borchers.
\newblock \emph{{\texttt{pracma}: Practical Numerical Math Functions}}.
\newblock R package version 2.2.5,
  \url{https://CRAN.R-project.org/package=pracma}, 2019.

\bibitem[Breslow(1974)]{breslow1974}
N.~Breslow.
\newblock {Covariance Analysis of Censored Survival Data}.
\newblock \emph{Biometrics}, 30\penalty0 (1):\penalty0 89--99, 1974.
\newblock \doi{10.2307/2529620}.

\bibitem[Brooks and Gelman(1998)]{brooks1998}
S.~P. Brooks and A.~Gelman.
\newblock {General Methods for Monitoring Convergence of Iterative
  Simulations}.
\newblock \emph{Journal of Computational and Graphical Statistics}, 7\penalty0
  (4):\penalty0 434--455, 1998.
\newblock \doi{10.1080/10618600.1998.10474787}.

\bibitem[Cai et~al.(2012)Cai, Zou, Peng, and Zhang]{smcure}
C.~Cai, Y.~Zou, Y.~Peng, and J.~Zhang.
\newblock \emph{{\texttt{smcure}: Fit Semiparametric Mixture Cure Models}}.
\newblock R package version 2.0,
  \url{https://CRAN.R-project.org/package=smcure}, 2012.

\bibitem[Carpenter et~al.(2017)Carpenter, Gelman, Hoffman, Lee, Goodrich,
  Betancourt, Brubaker, Guo, Li, and Riddell]{Stan}
B.~Carpenter, A.~Gelman, M.~Hoffman, D.~Lee, B.~Goodrich, M.~Betancourt,
  M.~Brubaker, J.~Guo, P.~Li, and A.~Riddell.
\newblock {Stan: A Probabilistic Programming Language}.
\newblock \emph{Journal of Statistical Software}, 76\penalty0 (1):\penalty0
  1--32, 2017.
\newblock \doi{10.18637/jss.v076.i01}.

\bibitem[Christensen et~al.(2011)Christensen, Wesley, Branscum, and
  Hanson]{christensen2011}
R.~Christensen, J.~Wesley, A.~Branscum, and T.~E. Hanson.
\newblock \emph{{Bayesian Ideas and Data Analysis: An Introduction for
  Scientists and Statisticians}}.
\newblock Chapman \& Hall/CRC, 1st edition, 2011.

\bibitem[Clayton(1978)]{clayton1978}
D.~G. Clayton.
\newblock {A Model for Association in Bivariate Life Tables and Its Application
  in Epidemiological Studies of Familial Tendency in Chronic Disease
  Incidence}.
\newblock \emph{Biometrika}, 65\penalty0 (1):\penalty0 141--151, 1978.
\newblock \doi{10.1093/biomet/65.1.141}.

\bibitem[Collett(2015)]{collet2015}
D.~Collett.
\newblock \emph{{Modelling Survival Data in Medical Research}}.
\newblock Chapman \& Hall/CRC, 3rd edition, 2015.

\bibitem[Cox(1972)]{cox1972}
D.~R. Cox.
\newblock {Regression Models and Life-Tables}.
\newblock \emph{Journal of the Royal Statistical Society. Series B
  (Methodological)}, 34\penalty0 (2):\penalty0 187--220, 1972.
\newblock \doi{10.1111/j.2517-6161.1972.tb00899.x}.

\bibitem[Crowley and Hu(1977)]{crowley1977}
J.~Crowley and M.~Hu.
\newblock {Covariance Analysis of Heart Transplant Survival Data}.
\newblock \emph{Journal of the American Statistical Association}, 72\penalty0
  (357):\penalty0 27--36, 1977.
\newblock \doi{10.1080/01621459.1977.10479903}.

\bibitem[{DeGruttola} and Tu(1994)]{gruttola1994}
V.~{DeGruttola} and X.~M. Tu.
\newblock {Modelling Progression of CD4-Lymphocyte Count and Its Relationship
  to Survival Time}.
\newblock \emph{Biometrics}, 50\penalty0 (4):\penalty0 1003--1014, 1994.
\newblock \doi{10.2307/2533439}.

\bibitem[Dettenkofer et~al.(2005)Dettenkofer, {Wenzler-Rottele}, Babikir,
  Bertz, Ebner, Meyer, Ruden, Gastmeier, and Daschner]{dettenkofer2005}
M.~Dettenkofer, S.~{Wenzler-Rottele}, R.~Babikir, H.~Bertz, W.~Ebner, E.~Meyer,
  H.~Ruden, P.~Gastmeier, and F.~D. Daschner.
\newblock {Surveillance of Nosocomial Sepsis and Pneumonia in Patients with a
  Bone Marrow or Peripheral Blood Stem Cell Transplant: A Multicenter Project}.
\newblock \emph{Clinical Infectious Diseases}, 40\penalty0 (7):\penalty0
  926--931, 2005.
\newblock \doi{10.1086/428046}.

\bibitem[Faucett and Thomas(1996)]{faucett1996}
C.~L. Faucett and D.~C. Thomas.
\newblock {Simultaneously Modelling Censored Survival Data and Repeatedly
  Measured Covariates: A Gibbs Sampling Approach}.
\newblock \emph{Statistics in Medicine}, 15\penalty0 (15):\penalty0 1663--1685,
  1996.
\newblock
  \doi{10.1002/(sici)1097-0258(19960815)15:15<1663::aid-sim294>3.0.co;2-1}.

\bibitem[Ge and Chen(2012)]{ge2012}
M.~Ge and M.~H. Chen.
\newblock {Bayesian Inference of the Fully Specified Subdistribution Model for
  Survival Data with Competing Risks}.
\newblock \emph{Lifetime Data Analysis}, 18\penalty0 (3):\penalty0 339--363,
  2012.
\newblock \doi{10.1007/s10985-012-9221-9}.

\bibitem[Gelman and Rubin(1992)]{gelman1992}
A.~Gelman and D.~B. Rubin.
\newblock {Inference from Iterative Simulation using Multiple Sequences}.
\newblock \emph{Statistical Science}, 7\penalty0 (4):\penalty0 457--472, 1992.
\newblock \doi{10.1214/ss/1177011136}.

\bibitem[Gelman et~al.(2013)Gelman, Carlin, Stern, Dunson, Vehtari, and
  Rubin]{bda2013}
A.~Gelman, J.~B. Carlin, H.~S. Stern, D.~B. Dunson, A.~Vehtari, and D.~B.
  Rubin.
\newblock \emph{{Bayesian Data Analysis}}.
\newblock Chapman \& Hall/CRC, 3rd edition, 2013.

\bibitem[Gilks et~al.(1994)Gilks, Thomas, and Spiegelhalter]{gilks1994}
W.~R. Gilks, A.~Thomas, and D.~J. Spiegelhalter.
\newblock {A Language and Program for Complex Bayesian Modelling}.
\newblock \emph{The Statistician}, 43\penalty0 (1):\penalty0 169--177, 1994.
\newblock \doi{10.2307/2348941}.

\bibitem[Grambauer and Neudecker(2011)]{compeir}
N.~Grambauer and A.~Neudecker.
\newblock \emph{{\texttt{compeir}: Event-Specific Incidence Rates for Competing
  Risks Data}}.
\newblock R package version 1.0,
  \url{https://CRAN.R-project.org/package=compeir}, 2011.

\bibitem[Ha et~al.(2018)Ha, Noh, Kim, and Lee]{frailtyHL}
I.~D. Ha, M.~Noh, J.~Kim, and Y.~Lee.
\newblock \emph{{\texttt{frailtyHL}: Frailty Models via Hierarchical
  Likelihood}}.
\newblock R package version 2.2,
  \url{https://CRAN.R-project.org/package=frailtyHL}, 2018.

\bibitem[Han et~al.(2014)Han, Yu, Dignamc, and Rathouzb]{han2014}
B.~Han, M.~Yu, J.~J. Dignamc, and P.~J. Rathouzb.
\newblock {Bayesian Approach for Flexible Modeling of Semicompeting Risks
  Data}.
\newblock \emph{Statistics in Medicine}, 33\penalty0 (29):\penalty0 5111--5125,
  2014.
\newblock \doi{10.1002/sim.6313}.

\bibitem[Hogan and Laird(1997)]{hogan1997}
J.~W. Hogan and N.~M. Laird.
\newblock {Mixture Models for the Joint Distribution of Repeated Measures and
  Event Times}.
\newblock \emph{Statistics in Medicine}, 16\penalty0 (3):\penalty0 239--257,
  1997.
\newblock
  \doi{10.1002/(sici)1097-0258(19970215)16:3<239::aid-sim483>3.0.co;2-x}.

\bibitem[Hogan and Laird(1998)]{hogan1998}
J.~W. Hogan and N.~M. Laird.
\newblock {Increasing Efficiency from Censored Survival Data by using Random
  Effects to Model Longitudinal Covariates}.
\newblock \emph{Statistical Methods in Medical Research}, 7\penalty0
  (1):\penalty0 28--48, 1998.
\newblock \doi{10.1177/096228029800700104}.

\bibitem[Hougaard(2000)]{hougaard2000}
P.~Hougaard.
\newblock \emph{{Analysis of Multivariate Survival Data}}.
\newblock Springer, 1st edition, 2000.

\bibitem[Ibrahim et~al.(2001)Ibrahim, Chen, and Sinha]{ibrahim2001}
J.~G. Ibrahim, M.~H. Chen, and D.~Sinha.
\newblock \emph{{Bayesian Survival Analysis}}.
\newblock Springer, 1st edition, 2001.

\bibitem[Jara et~al.(2018)Jara, Hanson, Quintana, Mueller, and
  Rosner]{DPpackage}
A.~Jara, T.~Hanson, F.~Quintana, P.~Mueller, and G.~Rosner.
\newblock \emph{{\texttt{DPpackage}: Bayesian Nonparametric Modeling in R}}.
\newblock R package version 1.1-7.4,
  \url{https://CRAN.R-project.org/package=DPpackage}, 2018.

\bibitem[Kalbfleisch and Prentice(2002)]{kalbfleisch2002}
J.~D. Kalbfleisch and R.~L. Prentice.
\newblock \emph{{The Statistical Analysis of Failure Time Data}}.
\newblock John Wiley \& Sons, 2nd edition, 2002.

\bibitem[Kardaun(1983)]{kardaun1983}
O.~Kardaun.
\newblock {Statistical Survival Analysis of Male Larynx-Cancer Patients - A
  Case Study}.
\newblock \emph{Statistica Neerlandica}, 37\penalty0 (3):\penalty0 103--125,
  1983.
\newblock \doi{10.1111/j.1467-9574.1983.tb00806.x}.

\bibitem[Kersey et~al.(1987)Kersey, Weisdorf, Nesbit, {LeBien}, Woods,
  {McGlave}, Kim, Vallera, Goldman, Bostrom, Hurd, and Ramsay]{kersey1987}
J.~H. Kersey, D.~Weisdorf, M.~E. Nesbit, T.~W. {LeBien}, W.~G. Woods, P.~B.
  {McGlave}, T.~Kim, D.~A. Vallera, A.~I. Goldman, B.~Bostrom, D.~Hurd, and
  N.~K.~C. Ramsay.
\newblock {Comparison of Autologous and Allogeneic Bone Marrow Transplantation
  for Treatment of High-Risk Refractory Acute Lymphoblastic Leukemia}.
\newblock \emph{New England Journal of Medicine}, 317\penalty0 (8):\penalty0
  461--467, 1987.
\newblock \doi{10.1056/nejm198708203170801}.

\bibitem[Klein and Moeschberger(2003)]{klein2003}
J.~P. Klein and M.~L. Moeschberger.
\newblock \emph{{Survival Analysis: Techniques for Censored and Truncated
  Data}}.
\newblock Springer, 2nd edition, 2003.

\bibitem[Klein et~al.(2012)Klein, Moeschberger, and Yan]{KMsurv}
J.~P. Klein, M.~L. Moeschberger, and J.~Yan.
\newblock \emph{{\texttt{KMsurv}: Data Sets from Klein and Moeschberger (1997),
  Survival Analysis}}.
\newblock R package version 0.1-5,
  \url{https://CRAN.R-project.org/package=KMsurv}, 2012.

\bibitem[Klein et~al.(2013)Klein, {van Houwelingen}, Ibrahim, and
  Scheike]{klein2013}
J.~P. Klein, H.~C. {van Houwelingen}, J.~G. Ibrahim, and T.~H. Scheike.
\newblock \emph{{Handbook of Survival Analysis}}.
\newblock Chapman \& Hall/CRC, 1st edition, 2013.

\bibitem[Kleinbaum and Klein(2012)]{kleinbaum2012}
D.~G. Kleinbaum and M.~Klein.
\newblock \emph{{Survival Analysis: A Self-Learning Text}}.
\newblock Springer, 3rd edition, 2012.

\bibitem[Kom{\'a}rek(2018)]{bayesSurv}
A.~Kom{\'a}rek.
\newblock \emph{{\texttt{bayesSurv}: Bayesian Survival Regression with Flexible
  Error and Random Effects Distributions}}.
\newblock R package version 3.2,
  \url{https://CRAN.R-project.org/package=bayesSurv}, 2018.

\bibitem[Laird and Ware(1982)]{laird1982}
N.~M. Laird and J.~H. Ware.
\newblock {Random-Effects Models for Longitudinal Data}.
\newblock \emph{Biometrics}, 38\penalty0 (4):\penalty0 963--974, 1982.
\newblock \doi{10.2307/2529876}.

\bibitem[L{\'a}zaro et~al.(2020)L{\'a}zaro, Armero, and Alvares]{lazaro2020}
E.~L{\'a}zaro, C.~Armero, and D.~Alvares.
\newblock {Bayesian Regularization for Flexible Baseline Hazard Functions in
  Cox Survival Models}.
\newblock \emph{Submitted}, 0\penalty0 (0):\penalty0 0--0, 2020.

\bibitem[Lee et~al.(2016)Lee, Dominici, Schrag, and Haneuse]{lee2016}
K.~H. Lee, F.~Dominici, D.~Schrag, and S.~Haneuse.
\newblock {Hierarchical Models for Semicompeting Risks Data with Application to
  Quality of End-of-life Care for Pancreatic Cancer}.
\newblock \emph{Journal of the American Statistical Association}, 111\penalty0
  (515):\penalty0 1075--1095, 2016.
\newblock \doi{10.1080/01621459.2016.1164052}.

\bibitem[Lee et~al.(2019)Lee, Lee, Alvares, and Haneuse]{SemiCompRisks}
K.~H. Lee, C.~Lee, D.~Alvares, and S.~Haneuse.
\newblock \emph{{\texttt{SemiCompRisks}: Hierarchical Models for Parametric and
  Semi-Parametric Analyses of Semi-Competing Risks Data}}.
\newblock R package version 3.2,
  \url{https://CRAN.R-project.org/package=SemiCompRisks}, 2019.

\bibitem[Lin et~al.(2015)Lin, Cai, Wang, and Zhang]{lin2015}
X.~Lin, B.~Cai, L.~Wang, and Z.~Zhang.
\newblock {A Bayesian Proportional Hazards Model for General Interval-Censored
  Data}.
\newblock \emph{Lifetime Data Analysis}, 21\penalty0 (3):\penalty0 470--490,
  2015.
\newblock \doi{10.1007/s10985-014-9305-9}.

\bibitem[Lunn et~al.(2012)Lunn, Jackson, Best, Thomas, and
  Spiegelhalter]{lunn2012}
D.~Lunn, C.~Jackson, N.~Best, A.~Thomas, and D.~Spiegelhalter.
\newblock \emph{{The \texttt{BUGS} Book: A Practical Introduction to Bayesian
  Analysis}}.
\newblock Chapman \& Hall/CRC, 1st edition, 2012.

\bibitem[{McGilchrist} and Aisbett(1991)]{mcgilchrist1991}
C.~A. {McGilchrist} and C.~W. Aisbett.
\newblock {Regression with Frailty in Survival Analysis}.
\newblock \emph{Biometrics}, 47\penalty0 (2):\penalty0 461--466, 1991.
\newblock \doi{10.2307/2532138}.

\bibitem[Meira-Machado and Roca-Pardinas(2012)]{p3state.msm}
L.~Meira-Machado and J.~Roca-Pardinas.
\newblock \emph{{\texttt{p3state.msm}: Analyzing Survival Data From
  Illness-Death Model}}.
\newblock R package version 1.3,
  \url{https://CRAN.R-project.org/package=p3state.msm}, 2012.

\bibitem[Mitra and M{\"u}ller(2015)]{mitra2015}
R.~Mitra and P.~M{\"u}ller, editors.
\newblock \emph{{Nonparametric Bayesian Inference in Biostatistics}}.
\newblock Springer, 1st edition, 2015.

\bibitem[Murray et~al.(2016)Murray, Hobbs, Sargent, and Carlin]{murray2016}
T.~A. Murray, B.~P. Hobbs, D.~J. Sargent, and B.~P. Carlin.
\newblock {Flexible Bayesian Survival Modeling with Semiparametric
  Time-dependent and Shape-restricted Covariate Effects}.
\newblock \emph{Bayesian Analysis}, 11\penalty0 (2):\penalty0 381--402, 2016.
\newblock \doi{10.1214/15-BA954}.

\bibitem[Ntzoufras(2009)]{ntzoufras2009}
I.~Ntzoufras.
\newblock \emph{{Bayesian Modeling using \texttt{WinBUGS}}}.
\newblock John Wiley \& Sons, 1st edition, 2009.

\bibitem[Pan et~al.(2017)Pan, Cai, Wang, and Lin]{ICBayes}
C.~Pan, B.~Cai, L.~Wang, and X.~Lin.
\newblock \emph{{\texttt{ICBayes}: Bayesian Semiparametric Models for
  Interval-Censored Data}}.
\newblock R package version 1.1,
  \url{https://CRAN.R-project.org/package=ICBayes}, 2017.

\bibitem[Plummer(2003)]{plummer2003}
M.~Plummer.
\newblock {\texttt{JAGS}: A Program for Analysis of Bayesian Graphical Models
  using Gibbs Sampling}.
\newblock pages 1--10. Proceedings of the 3rd International Workshop on
  Distributed Statistical Computing, Vienna, Austria, 2003.

\bibitem[Plummer(2018)]{rjags}
M.~Plummer.
\newblock \emph{{\texttt{rjags}: Bayesian Graphical Models Using MCMC}}.
\newblock R package version 4-8,
  \url{https://CRAN.R-project.org/package=rjags}, 2018.

\bibitem[Plummer et~al.(2019)Plummer, Best, Cowles, Vines, Sarkar, Bates,
  Almond, and Magnusson]{coda}
M.~Plummer, N.~Best, K.~Cowles, K.~Vines, D.~Sarkar, D.~Bates, R.~Almond, and
  A.~Magnusson.
\newblock \emph{{\texttt{coda}: Output Analysis and Diagnostics for MCMC}}.
\newblock R package version 0.19-3,
  \url{https://CRAN.R-project.org/package=coda}, 2019.

\bibitem[{Proust-Lima} et~al.(2014){Proust-Lima}, S{\'e}ne, Taylor, and
  {Jacqmin-Gadda}]{proust-Lima2014}
C.~{Proust-Lima}, M.~S{\'e}ne, J.~M.~G. Taylor, and H.~{Jacqmin-Gadda}.
\newblock {Joint Latent Class Models for Longitudinal and Time-To-Event Data: A
  Review}.
\newblock \emph{Statistical Methods in Medical Research}, 23\penalty0
  (1):\penalty0 74--90, 2014.
\newblock \doi{10.1177/0962280212445839}.

\bibitem[Putter et~al.(2007)Putter, Fiocco, and Geskus]{putter2007}
H.~Putter, M.~Fiocco, and R.~B. Geskus.
\newblock {Tutorial in Biostatistics: Competing Risks and Multi-State Models}.
\newblock \emph{Statistics in Medicine}, 26\penalty0 (11):\penalty0 2389--2430,
  2007.
\newblock \doi{10.1002/sim.2712}.

\bibitem[{R Core Team}(2020)]{R}
{R Core Team}.
\newblock \emph{{R: A Language and Environment for Statistical Computing}}.
\newblock R Foundation for Statistical Computing,
  \url{https://www.R-project.org}, 2020.

\bibitem[Raftery et~al.(2018)Raftery, Hoeting, Volinsky, Painter, and
  Yeung]{BMA}
A.~Raftery, J.~Hoeting, C.~Volinsky, I.~Painter, and K.~Y. Yeung.
\newblock \emph{{\texttt{BMA}: Bayesian Model Averaging}}.
\newblock R package version 3.18.9,
  \url{https://CRAN.R-project.org/package=BMA}, 2018.

\bibitem[Rizopoulos(2012)]{rizopoulos2012}
D.~Rizopoulos.
\newblock \emph{{Joint Models for Longitudinal and Time-To-Event Data: With
  Applications in R}}.
\newblock Chapman \& Hall/CRC, 1st edition, 2012.

\bibitem[Rizopoulos(2019)]{JMbayes}
D.~Rizopoulos.
\newblock \emph{{\texttt{JMbayes}: Joint Modeling of Longitudinal and
  Time-to-Event Data Under a Bayesian Approach}}.
\newblock R package version 0.8-83,
  \url{https://CRAN.R-project.org/package=JMbayes}, 2019.

\bibitem[Rizopoulos et~al.(2015)Rizopoulos, Taylor, Rosmalen, Steyerberg, and
  Takkenberg]{rizopoulos2015}
D.~Rizopoulos, J.~M.~G. Taylor, J.~V. Rosmalen, E.~W. Steyerberg, and J.~J.~M.
  Takkenberg.
\newblock {Personalized Screening Intervals for Biomarkers using Joint Models
  for Longitudinal and Survival Data}.
\newblock \emph{Biostatistics}, 17\penalty0 (1):\penalty0 149--164, 2015.
\newblock \doi{10.1093/biostatistics/kxv031}.

\bibitem[Royston(2011)]{royston2011}
P.~Royston.
\newblock {Estimating a Smooth Baseline Hazard Function for the Cox Model}.
\newblock {R}esearch report 314, MRC Clinical Trials Unit and University
  College London, London, 2011.

\bibitem[Rue et~al.(2009)Rue, Martino, and Chopin]{INLA}
H.~Rue, S.~Martino, and N.~Chopin.
\newblock {Approximate Bayesian Inference for Latent Gaussian Models Using
  Integrated Nested Laplace Approximations}.
\newblock \emph{Journal of the Royal Statistical Society. Series B
  (Methodological)}, 71\penalty0 (2):\penalty0 319--392, 2009.
\newblock \doi{10.1111/j.1467-9868.2008.00700.x}.

\bibitem[Ru{\'e} et~al.(2017)Ru{\'e}, Andrinopoulou, Alvares, Armero, Forte,
  and Blanch]{rue2017}
M.~Ru{\'e}, E.~R. Andrinopoulou, D.~Alvares, C.~Armero, A.~Forte, and
  L.~Blanch.
\newblock {Bayesian Joint Modeling of Bivariate Longitudinal and Competing
  Risks Data: An Application to Study Patient-Ventilator Asynchronies in
  Critical Care Patients}.
\newblock \emph{Biometrical Journal}, 59\penalty0 (6):\penalty0 1184--1203,
  2017.
\newblock \doi{10.1002/bimj.201600221}.

\bibitem[Sahu et~al.(1997)Sahu, Dey, Aslanidou, and Sinha]{sahu1997}
S.~K. Sahu, D.~K. Dey, H.~Aslanidou, and D.~Sinha.
\newblock {A Weibull Regression Model with Gamma Frailties for Multivariate
  Survival Data}.
\newblock \emph{Lifetime Data Analysis}, 3\penalty0 (2):\penalty0 123--137,
  1997.
\newblock \doi{10.1023/a:1009605117713}.

\bibitem[Schuurman et~al.(2016)Schuurman, Grasman, and Hamaker]{schuurman2016}
N.~K. Schuurman, R.~P. P.~P. Grasman, and E.~L. Hamaker.
\newblock {A Comparison of Inverse-Wishart Prior Specifications for Covariance
  Matrices in Multilevel Autoregressive Models}.
\newblock \emph{Multivariate Behavioral Research}, 51\penalty0 (2-3):\penalty0
  185--206, 2016.
\newblock \doi{10.1080/00273171.2015.1065398}.

\bibitem[Sharabiani and Mahani(2019)]{CFC}
M.~T.~A. Sharabiani and A.~S. Mahani.
\newblock \emph{{\texttt{CFC}: Cause-Specific Framework for Competing-Risk
  Analysis}}.
\newblock R package version 1.1.2,
  \url{https://CRAN.R-project.org/package=CFC}, 2019.

\bibitem[Smyth et~al.(2020)Smyth, Hu, Dunn, Phipson, and Chen]{statmod}
G.~Smyth, Y.~Hu, P.~Dunn, B.~Phipson, and Y.~Chen.
\newblock \emph{{\texttt{statmod}: Statistical Modeling}}.
\newblock R package version 1.4.34,
  \url{https://CRAN.R-project.org/package=statmod}, 2020.

\bibitem[{Stan Development Team}(2020)]{RStan}
{Stan Development Team}.
\newblock \emph{{RStan: the R Interface to Stan}}.
\newblock Stan, \url{http://mc-stan.org/}, 2020.

\bibitem[Taylor et~al.(2018)Taylor, Rowlingson, and Zheng]{spatsurv}
B.~M. Taylor, B.~S. Rowlingson, and Z.~Zheng.
\newblock \emph{{\texttt{spatsurv}: Bayesian Spatial Survival Analysis with
  Parametric Proportional Hazards Models}}.
\newblock R package version 1.2,
  \url{https://CRAN.R-project.org/package=spatsurv}, 2018.

\bibitem[Taylor et~al.(2013)Taylor, Park, Ankerst, {Proust-Lima}, Williams,
  Kestin, Bae, Pickles, and Sandler]{taylor2013}
J.~M.~G. Taylor, Y.~Park, D.~P. Ankerst, C.~{Proust-Lima}, S.~Williams,
  L.~Kestin, K.~Bae, T.~Pickles, and H.~Sandler.
\newblock {Real-Time Individual Predictions of Prostate Cancer Recurrence using
  Joint Models}.
\newblock \emph{Biometrics}, 69\penalty0 (1):\penalty0 206--213, 2013.
\newblock \doi{10.1111/j.1541-0420.2012.01823.x}.

\bibitem[Tsiatis et~al.(1995)Tsiatis, {DeGruttola}, and Wulfsohn]{tsiatis1995}
A.~A. Tsiatis, V.~{DeGruttola}, and M.~S. Wulfsohn.
\newblock {Modeling the Relationship of Survival to Longitudinal Data Measured
  with Error. Applications to Survival and CD4 Counts in Patients with AIDS}.
\newblock \emph{Journal of the American Statistical Association}, 90\penalty0
  (429):\penalty0 27--37, 1995.
\newblock \doi{10.1080/01621459.1995.10476485}.

\bibitem[Umlauf et~al.(2019)Umlauf, Kneib, and Klein]{BayesX}
N.~Umlauf, T.~Kneib, and N.~Klein.
\newblock \emph{{\texttt{BayesX}: R Utilities Accompanying the Software Package
  BayesX}}.
\newblock R package version 0.3-1,
  \url{https://CRAN.R-project.org/package=BayesX}, 2019.

\bibitem[Vaupel et~al.(1979)Vaupel, Manton, and Stallard]{vaupel1979}
J.~W. Vaupel, K.~G. Manton, and E.~Stallard.
\newblock {The Impact of Heterogeneity in Individual Frailty on the Dynamics of
  Mortality}.
\newblock \emph{Demography}, 16\penalty0 (3):\penalty0 439--454, 1979.
\newblock \doi{10.2307/2061224}.

\bibitem[Wang et~al.(2017)Wang, Chen, Wang, and Yan]{dynsurv}
X.~Wang, M.~H. Chen, W.~Wang, and J.~Yan.
\newblock \emph{{\texttt{dynsurv}: Dynamic Models for Survival Data}}.
\newblock R package version 0.3-6,
  \url{https://CRAN.R-project.org/package=dynsurv}, 2017.

\bibitem[Wu and Carroll(1988)]{wu1988}
M.~C. Wu and R.~J. Carroll.
\newblock {Estimation and Comparison of Changes in the Presence of Informative
  Right Censoring by Modeling the Censoring Process}.
\newblock \emph{Biometrics}, 44\penalty0 (1):\penalty0 175--188, 1988.
\newblock \doi{10.2307/2531905}.

\bibitem[Wulfsohn and Tsiatis(1997)]{wulfsohn1997}
M.~S. Wulfsohn and A.~A. Tsiatis.
\newblock {A Joint Model for Survival and Longitudinal Data Measured with
  Error}.
\newblock \emph{Biometrics}, 53\penalty0 (1):\penalty0 330--339, 1997.
\newblock \doi{10.2307/2533118}.

\bibitem[Zhou and Hanson(2018)]{spBayesSurv}
H.~Zhou and T.~Hanson.
\newblock \emph{{\texttt{spBayesSurv}: Bayesian Modeling and Analysis of
  Spatially Correlated Survival Data}}.
\newblock R package version 1.1.3,
  \url{https://CRAN.R-project.org/package=spBayesSurv}, 2018.

\end{thebibliography}

\appendix

\section{Implementation of the transition probabilities of the semi-Markov multi-state model based on posterior samples} \label{app1}

Figure~\ref{fig:ms} was generated from the transition probabilities \eqref{eq:tp11}--\eqref{eq:tp33} based on posterior samples of the illness-death model parameter (\ref{eq:scr1})--(\ref{eq:scr3}). In particular, we set a grid of time values to evaluate the transition probabilities and used the median values of \texttt{age} and \texttt{year} covariates. In this example, we are interested in the transition probabilities for patients with and without prior bypass surgery (i.e., \texttt{surgery} equal to 1 or 0, respectively):
\begin{verbatim}
R> grid <- seq(0, 1000, length.out = 51)
R> x0 <- c(median(heart2$age), median(heart2$year), 0)
R> x1 <- c(median(heart2$age), median(heart2$year), 1)
\end{verbatim}

For our Weibull baseline hazard specification, the transition probability $p_{11}(s, t \mid \boldsymbol \theta)$ (see Equation~\ref{eq:tp11}) can be calculated analytically and is expressed by:
\begin{equation}
p_{11}(s, t \mid \bm \theta) = \exp\Big\{-\lambda_{1} \eta_{1} \big[t^{\alpha_{1}} - s^{\alpha_{1}}\big] -\lambda_{3} \eta_{3} \big[t^{\alpha_{3}} - s^{\alpha_{3}}\big]\Big\}, \label{weibullp11}
\end{equation}

\noindent where $\eta_{k}=\exp\big({\bm x}^{\top}{\bm \beta}_{k}\big)$, for $k=1,3$. Equation~(\ref{weibullp11}) is implemented as follows:
\begin{verbatim}
R> p11_s_t <- function(tt, ss, l1, a1, b1, l2, a2, b2, x){
+    H1 <- l1 * exp(sum(b1 * x)) * (tt^a1 - ss^a1)
+    H2 <- l2 * exp(sum(b2 * x)) * (tt^a2 - ss^a2)
+    return(exp(-H1 - H2))
+  }
\end{verbatim}

Next, we have created a function \texttt{mcmc\_p11\_s\_t} that takes posterior samples of each parameter from \texttt{JAGS} and computes $p_{11}(s, t \mid \boldsymbol \theta)$ in (\ref{weibullp11}) for a grid of time values (variable \texttt{grid}) using covariates \texttt{x} (here \texttt{x0} or \texttt{x1}). Note that \texttt{mcmc\_p11\_s\_t} is based on the \texttt{mclapply} function, available from the \texttt{parallel} package to speed computations up \citep{R}.
\begin{verbatim}
R> library("parallel")
R> options(mc.cores = detectCores())
R> mcmc_p11_s_t <- function(t.pred, s.pred, l1, a1, b1, l2, a2, b2, x){
+    # Sub-sample to speed up computations
+    samples.idx <- sample(1:length(l1), 200)
+    res <- sapply(t.pred, function(tt){
+      aux <- mclapply(samples.idx, function(i){
+        p11_s_t(tt, ss = s.pred, l1 = l1[i], a1 = a1[i], b1 = b1[i,],
+          l2 = l2[i], a2 = a2[i], b2 = b2[i,], x = x)
+        })
+        return(mean(unlist(aux)))
+    })
+    return(res)
+  }
\end{verbatim}

As previously mentioned, we are interested in the transition probabilities for patients with and without prior bypass surgery. In particular, $p_{11}(s, t \mid \boldsymbol \theta)$ for $s=0$ can be calculated as follows:
\begin{verbatim}
R> p11_0_t_surgery0 <- mcmc_p11_s_t(t.pred = grid, s.pred = 0, l1 = lambda1, a1 = alpha1,
+    b1 = cbind(beta11, beta21, beta31), l2 = lambda2, a2 = alpha2,
+    b2 = cbind(beta12, beta22, beta32), x = x0) 
R> p11_0_t_surgery1 <- mcmc_p11_s_t(t.pred = grid, s.pred = 0, l1 = lambda1, a1 = alpha1,
+    b1 = cbind(beta11, beta21, beta31), l2 = lambda2, a2 = alpha2,
+    b2 = cbind(beta12, beta22, beta32), x = x1)
\end{verbatim}

Finally, these transition probabilities for \texttt{x0} and \texttt{x1} are plotted in Figure~\ref{fig:ms} using the following code:
\begin{verbatim}
R> library("ggplot2")
R> surg.colour <- rep(c("no", "yes"), each = length(grid))
R> df11 <- data.frame(time = rep(grid, 2), surgery = factor(surg.colour),
+    transition = c(p11_0_t_surgery0, p11_0_t_surgery1))
R> p11 <- ggplot(data = df11, aes(x = time, y = transition, group = surgery, colour = surgery)) + 
+    geom_line() + annotate("text", x = 880, y = 1, label=expression(1 %->% 1), size=5) + 
+    ylab("probability") + theme(legend.position = "top") + ylim(0,1) + theme_bw()
\end{verbatim}

The transition probability $p_{22}(s, t \mid \boldsymbol \theta)$ (see Equation~\ref{eq:tp22}) using the Weibull baseline hazard specification is written as:
\begin{equation}
p_{22}(s, t \mid \bm \theta) = \frac{\lambda_{1}\alpha_{1}\eta_{1}}{1-\exp\big\{-\lambda_{1}\eta_{1}s^{\alpha_{1}}\big\}}\int_{0}^{s}u^{\alpha_{1}-1}\exp\Big\{-\lambda_{1} \eta_{1} u^{\alpha_{1}} - \lambda_{3} \eta_{3} \big[(t-u)^{\alpha_{3}} - (s-u)^{\alpha_{3}}\big]\Big\}\,\mbox{d}u, \label{weibullp22}
\end{equation} 

\noindent where $\eta_{k}=\exp\big({\bm x}^{\top}{\bm \beta}_{k}\big)$, for $k=1,3$. The integral in (\ref{weibullp22}) has no closed form, so some approximate method of integration is required. To do this, we first have created a function \texttt{int\_p22} to calculate this integral:
\begin{verbatim}
R> int_p22 <- function(u, tt, ss, l1, a1, b1, l3, a3, b3, x){
+    F1 <- 1 - exp(-l1 * exp(sum(b1 * x)) * ss^a1)
+    u1 <- ifelse(u > 0, u^(a1 - 1),0)
+    h1 <- l1 * a1 * exp(sum(b1 * x)) * u1
+    H1 <- l1 * exp(sum(b1 * x)) * u^a1
+    H3 <- l3 * exp(sum(b3 * x)) * ((tt - u)^a3 - (ss - u)^a3)
+    return(h1 * exp(-H1 - H3) / F1)
+  }
\end{verbatim}

Next, we have created a function \texttt{p22\_s\_t} that computes $p_{22}(s, t \mid \boldsymbol \theta)$ in (\ref{weibullp22}) by integrating out the \texttt{int\_p22} function using the \texttt{integral} function available from the \texttt{pracma} package \citep{pracma}.
\begin{verbatim}
R> library("pracma")
R> p22_s_t <- function(tt, ss, l1, a1, b1, l3, a3, b3, x){
+    return(integral(int_p22, xmin = 0, xmax = ss, method = "Kronrod", tt = tt,
+      ss = ss, l1 = l1, a1 = a1, b1 = b1, l3 = l3, a3 = a3, b3 = b3, x = x))
+  }
\end{verbatim}

Analogous to the previous case, we have created a function \texttt{mcmc\_p22\_s\_t} that takes posterior samples of each parameter from \texttt{JAGS} and computes $p_{22}(s, t \mid \boldsymbol \theta)$ in (\ref{weibullp22}) for a grid of time values (variable \texttt{grid}) using covariates \texttt{x} (here \texttt{x0} or \texttt{x1}).
\begin{verbatim}
R> mcmc_p22_s_t <- function(t.pred, s.pred, l1, a1, b1, l3, a3, b3, x){
+    # Sub-sample to speed up computations
+    samples.idx <- sample(1:length(l1), 200)
+    res <- sapply(t.pred, function(tt){
+      aux <- mclapply(samples.idx, function(i){
+        p22_s_t(tt, ss = s.pred, l1 = l1[i], a1 = a1[i], b1 = b1[i,],
+          l3 = l3[i], a3 = a3[i], b3 = b3[i,], x = x)
+      })
+      return(mean(unlist(aux)))
+    })
+    return(res)
+  }
\end{verbatim}

Note that the transition time from state 1 to state 2 prevents the case $s=0$ in $p_{22}(s, t \mid \boldsymbol \theta)$. To realistically get around this problem, we have set $s$ as the median time of patients who transitioned from state 1 to 2 ($s=26$) and add this amount to the grid of time values. So, we have calculated the transition probabilities (\ref{weibullp22}) for \texttt{x0} and \texttt{x1}:
\begin{verbatim}
R> m <- median(heart2$times1[heart2$delta == 1])
R> grid2 <- grid + m
R> p22_m_t_surgery0 <- mcmc_p22_s_t(t.pred = grid2, s.pred = m, l1 = lambda1, a1 = alpha1,
+    b1 = cbind(beta11, beta21, beta31), l3 = lambda3, a3 = alpha3,
+    b3 = cbind(beta13, beta23, beta33), x = x0) 
R> p22_m_t_surgery1 <- mcmc_p22_s_t(t.pred = grid2, s.pred = m, l1 = lambda1, a1 = alpha1, 
+    b1 = cbind(beta11, beta21, beta31), l3 = lambda3, a3 = alpha3, 
+    b3 = cbind(beta13, beta23, beta33), x = x1)
\end{verbatim}

Finally, these transition probabilities are plotted in Figure~\ref{fig:ms} using the following code:
\begin{verbatim}
R> df22 <- data.frame(time = rep(grid, 2), surgery = factor(surg.colour),
+    transition = c(p22_m_t_surgery0, p22_m_t_surgery1))
R> p22 <- ggplot(data = df22, aes(x = time, y = transition, group = surgery, colour = surgery)) + 
+    geom_line() + annotate("text", x = 880, y = 1, label=expression(2 %->% 2), size=5) + 
+    ylab("probability") + theme(legend.position = "top") + ylim(0,1) + theme_bw()
\end{verbatim}

The transition probability $p_{12}(s, t \mid \boldsymbol \theta)$ (see Equation~\ref{eq:tp12}) using the Weibull baseline hazard specification is written as:
\begin{equation}
p_{12}(s, t \mid \bm \theta) = \lambda_{1}\alpha_{1}\eta_{1}\int_{s}^{t}u^{\alpha_{1}-1}\exp\Big\{-\lambda_{1} \eta_{1} \big[u^{\alpha_{1}}-s^{\alpha_{1}}\big] -\lambda_{2} \eta_{2} \big[u^{\alpha_{2}}-s^{\alpha_{2}}\big] - \lambda_{3} \eta_{3} (t-u)^{\alpha_{3}}\Big\}\,\mbox{d}u, \label{weibullp12}
\end{equation} 

\noindent where $\eta_{k}=\exp\big({\bm x}^{\top}{\bm \beta}_{k}\big)$, for $k=1,2,3$. Analogous to the previous case, the integral in (\ref{weibullp12}) has no closed form, so some approximate method of integration is required. To do this, we first have created a function \texttt{int\_p12} to calculate this integral:
\begin{verbatim}
R> int_p12 <- function(u, tt, ss, l1, a1, b1, l2, a2, b2, l3, a3, b3, x){
+    u1 <- ifelse(u > 0, u^(a1 - 1), 0)
+    h1 <- l1 * a1 * exp(sum(b1 * x)) * u1
+    H1 <- l1 * exp(sum(b1 * x)) * (u^a1 - ss^a1)
+    H2 <- l2 * exp(sum(b2 * x)) * (u^a2 - ss^a2)
+    H3 <- l3 * exp(sum(b3 * x)) * (tt - u)^a3
+    return(h1 * exp(-H1 - H2 - H3))
+  }
\end{verbatim}

Next, we have created a function \texttt{p12\_s\_t} that computes $p_{12}(s, t \mid \boldsymbol \theta)$ in (\ref{weibullp12}) by integrating out the \texttt{int\_p12} function.
\begin{verbatim}
R> p12_s_t <- function(tt, ss, l1, a1, b1, l2, a2, b2, l3, a3, b3, x){
+    return(integral(int_p12, xmin = ss, xmax = tt, method = "Kronrod", tt = tt, 
+      ss = ss, l1 = l1, a1 = a1, b1 = b1, l2 = l2, a2 = a2, b2 = b2,
+      l3 = l3, a3 = a3, b3 = b3, x = x))
+  }
\end{verbatim}

As a last step, we have created a function \texttt{mcmc\_p12\_s\_t} that takes posterior samples of each parameter from \texttt{JAGS} and computes $p_{12}(s, t \mid \boldsymbol \theta)$ in (\ref{weibullp12}) for a grid of time values (variable \texttt{grid}) using covariates \texttt{x} (here \texttt{x0} or \texttt{x1}).
\begin{verbatim}
R> mcmc_p12_s_t <- function(t.pred, s.pred, l1, a1, b1, l2, a2, b2, l3, a3, b3, x){
+    # Sub-sample to speed up computations
+    samples.idx <- sample(1:length(l1), 200)
+    res <- sapply(t.pred, function(tt){
+      aux <- mclapply(samples.idx, function(i){
+        p12_s_t(tt, ss = s.pred, l1 = l1[i], a1 = a1[i], b1 = b1[i,], l2 = l2[i],
+          a2 = a2[i], b2 = b2[i,], l3 = l3[i], a3 = a3[i], b3 = b3[i,], x = x)
+      })
+      return(mean(unlist(aux)))
+    })
+    return(res)
+  }
\end{verbatim}

The transition probabilities $p_{12}(s, t \mid \boldsymbol \theta)$ of \texttt{x0} and \texttt{x1} for $s=0$ can be calculated as follows:
\begin{verbatim}
R> p12_0_t_surgery0 <- mcmc_p12_s_t(t.pred = grid, s.pred = 0, l1 = lambda1, a1 = alpha1, 
+    b1 = cbind(beta11, beta21, beta31), l2 = lambda2, a2 = alpha2, 
+    b2 = cbind(beta12, beta22, beta32), l3 = lambda3, a3 = alpha3, 
+    b3 = cbind(beta13, beta23, beta33), x = x0) 
R> p12_0_t_surgery1 <- mcmc_p12_s_t(t.pred = grid, s.pred = 0, l1 = lambda1, a1 = alpha1, 
+    b1 = cbind(beta11, beta21, beta31), l2 = lambda2, a2 = alpha2, 
+    b2 = cbind(beta12, beta22, beta32), l3 = lambda3, a3 = alpha3, 
+    b3 = cbind(beta13, beta23, beta33), x = x1)
\end{verbatim}

Finally, these transition probabilities are plotted in Figure~\ref{fig:ms} using the following code:
\begin{verbatim}
R> df12 <- data.frame(time = rep(grid, 2), surgery = factor(surg.colour),
+    transition = c(p12_0_t_surgery0, p12_0_t_surgery1))
R> p12 <- ggplot(data = df12, aes(x = time, y = transition, group = surgery, colour = surgery)) + 
+    geom_line() + annotate("text", x = 880, y = 1, label=expression(1 %->% 2), size=5) + 
+    ylab("probability") + theme(legend.position = "top") + ylim(0,1) + theme_bw()
\end{verbatim}

As shown in \eqref{eq:tp13} and \eqref{eq:tp23}, $p_{13}(s, t \mid \bm \theta)$ and $p_{23}(s, t \mid \bm \theta)$ are derived from the previously calculated transition probabilities. So, we can get them for \texttt{x0} and \texttt{x1} as follows:
\begin{verbatim}
R> p13_0_t_surgery0 <- 1 - p11_0_t_surgery0 - p12_0_t_surgery0
R> p13_0_t_surgery1 <- 1 - p11_0_t_surgery1 - p12_0_t_surgery1
R> p23_m_t_surgery0 <- 1 - p22_m_t_surgery0
R> p23_m_t_surgery1 <- 1 - p22_m_t_surgery1
\end{verbatim}

These transition probabilities are plotted in Figure~\ref{fig:ms} using the following code:
\begin{verbatim}
R> df13 <- data.frame(time = rep(grid, 2), surgery = factor(surg.colour),
+    transition = c(p13_0_t_surgery0, p13_0_t_surgery1))
R> df23 <- data.frame(time = rep(grid, 2), surgery = factor(surg.colour),
+    transition = c(p23_m_t_surgery0, p23_m_t_surgery1))
R> p13 <- ggplot(data = df13, aes(x = time, y = transition, group = surgery, colour = surgery)) + 
+    geom_line() + annotate("text", x = 880, y = 1, label=expression(1 %->% 3), size=5) + 
+    ylab("probability") + theme(legend.position = "top") + ylim(0,1) + theme_bw()
R> p23 <- ggplot(data = df23, aes(x = time, y = transition, group = surgery, colour = surgery)) + 
+    geom_line() + annotate("text", x = 880, y = 1, label=expression(2 %->% 3), size=5) + 
+    ylab("probability") + theme(legend.position = "top") + ylim(0,1) + theme_bw()
\end{verbatim}

To show these graphs in two rows and three columns, as in Figure~\ref{fig:ms}, we have used the \texttt{grid.arrange} function available from the \texttt{gridExtra} package \citep{gridextra}.
\begin{verbatim}
R> library("gridExtra")
R> grid.arrange(p11, p12, p13, p22, p23, nrow = 2, ncol = 3, 
+    layout_matrix = rbind(c(1, 2, 3), c(NA, 5, 6)))
\end{verbatim}

\end{document}